\PassOptionsToPackage{pdfpagelabels=false}{hyperref}
\PassOptionsToPackage{usenames, dvipsnames}{color}
\documentclass[useAMS,usenatbib]{mnras}

\usepackage{natbib}

\usepackage{graphicx}
\usepackage[fleqn]{amsmath}
\usepackage{threeparttable}
\usepackage{ulem}
\usepackage{pdflscape}
\usepackage{cleveref}
\usepackage{enumitem}
\usepackage[UKenglish]{babel}
\usepackage[UKenglish]{isodate}
\usepackage[usenames,dvipsnames]{color}
\usepackage{multicol}
\usepackage{stackengine}

\setlist[itemize,1]{labelsep=4pt,leftmargin=2pt,labelwidth=3pt, itemindent=5pt}

\defcitealias{Hopwood15}{Hop15}
\defcitealias{FFredshift}{Paper~II}
\defcitealias{FFlineID}{Paper~III}
\defcitealias{FFncc}{Paper~IV}

\include{epsf}

\newcommand{\herschel}{\textit{Herschel}}

\title[SPIRE FTS Spectral Feature Finder]{The \herschel\ SPIRE Fourier Transform Spectrometer Spectral Feature Finder I. The Spectral Feature Finder and Catalogue\thanks{\herschel\ was a European Space Agency (ESA) space observatory with science instruments provided by European-led Principal Investigator consortia and with important participation from NASA.}}

\author[R. Hopwood et al.]{R. Hopwood$^{1,2}$, 
I. Valtchanov$^{1}$, 
L. D. Spencer$^{3}$\thanks{E-mail: Locke.Spencer@uLeth.ca},
J. Scott$^{3}$,
C. Benson$^{3}$,
\newauthor
N. Marchili$^{4}$,
N. H\l{}adczuk$^{5,6}$,
E. T. Polehampton$^{7}$, 
N. Lu$^{8}$,
G. Makiwa$^{3}$,
\newauthor
D. A. Naylor$^{3}$,
B. G. Gom$^{3}$,
G. Noble$^{9}$,
and M. J. Griffin$^{10}$
\\
$^{1}$Telespazio Vega UK for ESA, European Space Astronomy Centre, Operations Department, 28691 Villanueva de la Ca\~nada, Spain\\
$^{2}$Department of Physics, Imperial College London, Prince Consort Road, London SW7 2AZ, UK \\
$^3$Department of Physics \& Astronomy, University of Lethbridge, 
Lethbridge, Alberta, T1K 3M4, Canada \\
$^{4}$
Istituto di Radioastronomia - INAF, Via Piero Gobetti 101, 40129, Bologna, Italy \\
$^{5}$ European Space Agency, ESAC, Camino Bajo del Castillo, 28692, Villanueva de la Ca$\tilde{n}$ada, Madrid, Spain \\
$^{6}$ Gran TeCan, S.A., Instituto de Astrof\'{i}sica de Canarias, C/ V\'{i}a L\'{a}ctea, S/N, 38205\\ 
 - San Crist\'{o}bal de La Laguna, S/C de Tenerife, Spain\\
$^{7}$RAL Space, Rutherford Appleton Laboratory, Chilton, Didcot, Oxfordshire, OX11 0QX, UK \\
$^{8}$South American Center for Astronomy, CAS, Universidad de Chile, Camino El Observatorio \#1515, Las Condes, Santiago, Chile\\
$^{9}$Department of Physics, McGill University, 3600 Rue University, Montreal, Quebec, H3A 2T8, Canada\\
$^{10}$School of Physics and Astronomy, Cardiff University, The Parade, Cardiff, CF24  3AA, UK\\
}

\def\doubleline{\vskip3pt\hrule\vskip1.5pt\hrule\vskip5pt}

\begin{document}

\date{Accepted 2020 June 03. Received 2020 June 03; in original form 2020 February 27}

\pagerange{\pageref{firstpage}--\pageref{lastpage}} \pubyear{2020}

\maketitle

\label{firstpage}

\begin{abstract}
We provide a detailed description of the \herschel /SPIRE Fourier Transform Spectrometer (FTS) Spectral Feature Finder (FF). The FF is an automated process designed to extract significant spectral features from SPIRE FTS data products. Optimising the number of features found in SPIRE-FTS spectra is challenging. The wide SPIRE-FTS frequency range ($447$–-$1568$\,GHz) leads to many molecular species and atomic fine structure lines falling within the observed bands. As the best spectral resolution of the SPIRE-FTS is $\sim$1.2\,GHz, there can be significant line blending, depending on the source type. In order to find, both efficiently and reliably, features in spectra associated with a wide range of sources, the FF iteratively searches for peaks over a number of signal-to-noise ratio (SNR) thresholds. For each threshold, newly identified features are rigorously checked before being added to the fitting model. At the end of each iteration, the FF simultaneously fits the continuum and features found, with the resulting residual spectrum used in the next iteration. The final FF products report the frequency of the features found and the associated SNRs. Line flux determination is not included as part of the FF products, as extracting reliable line flux from SPIRE-FTS data is a complex process that requires careful evaluation and analysis of the spectra on a case-by-case basis. The FF results are 100\% complete for features with SNR greater than 10 and 50--70\% complete at SNR of 5. The FF code and all FF products are publicly available via the \herschel\ Science Archive.
\end{abstract}

%
\section{Introduction}\label{sec:intro}
\vspace{-12pt}
The ESA \herschel\ Space Observatory \citep[\herschel;][]{pilbratt10} observed the far infrared and sub-millimetre (submm) sky from an orbit around the Sun-Earth L2 Lagrangian point, from May 2009 to April 2013. After an initial six-months of commissioning and performance demonstration, \herschel\ entered its routine operation phase, during which science and calibration observations were conducted. The Spectral and Photometric Imaging REceiver \citep[SPIRE;][]{Griffin10} was one of three focal plane instruments on board \herschel. SPIRE consisted of an imaging photometric camera and an imaging Fourier Transform Spectrometer (FTS). The SPIRE FTS's two bolometer arrays, Spectrometer Long Wavelength -- SLW (447--990\,GHz) and Spectrometer Short Wavelength -- SSW (958--1546\,GHz), provided simultaneous frequency coverage at high (HR; $\sim$1.2\,GHz) and low (LR; $\sim$25\,GHz) spectral resolution, across a wide frequency band in the submm. The bolometric detectors \citep{Turner2001} operated at $\sim$300\,mK with hexagonally close-packed feedhorn-coupled focal plane optics, providing sparse spatial sampling over an unvignetted field of view of 2$^\prime$ \citep{Dohlen2000}. 
%
%
%
%
Details on the SPIRE FTS data processing and data products are provided in the SPIRE Data Reduction Guide (SDRG), available in the \textit{Herschel} Explanatory Legacy Library.\footnote{The \herschel\ Explanatory Legacy Library contains all relevant documents for the three \herschel\ instruments, including SPIRE, available at the following URL: \url{https://www.cosmos.esa.int/web/herschel/legacy-documentation}} The calibration scheme and the calibration accuracy for the SPIRE FTS are described by \citet{Swinyard2014}, \citet{Marchili_HLR_MNRAS}
and \citet{Hopwood15}. Due to repeated referencing of \citet{Hopwood15}, the abbreviation \citetalias{Hopwood15} is used in further text within this paper.

This paper details the SPIRE FTS Spectral Feature Finder (FF) and the associated products, which are available in the \href{http://archives.esac.esa.int/hsa/whsa/}{\herschel\ Science Archive (HSA)}\footnote{\herschel\ Science Archive: \href{http://archives.esac.esa.int/hsa/whsa}{archives.esac.esa.int/hsa/whsa}}. The \href{https://doi.org/10.5270/esa-lysf2yi}{FF}\footnote{The \herschel\ SPIRE Spectral Feature Catalogue has been assigned an ESA Digital Object Identifier (DOI) and is available at: \href{https://doi.org/10.5270/esa-lysf2yi}{doi.org/10.5270/esa-lysf2yi}.} is an automated process that finds and fits significant spectral features in all SPIRE FTS observations, and, where appropriate, produces catalogue entries associated with the features found. The main motivation behind developing the FF was as a data mining aid for the HSA. To this end, the FF products provide an initial data analysis shortcut, by allowing quick inspection of one or multiple FTS observations and facilitating the search for spectral features of interest in all SPIRE FTS observations. Use of the FF products should not, however, replace the detailed data inspection and hands-on spectral analysis necessary to produce conclusive scientific results.
There are three companion papers which discuss aspects of the FF: a radial source velocity paper 
\citep[\citetalias{FFredshift}:][see also \citealt{ScottZ16}]{FFredshift}, a line identification paper \citep[\citetalias{FFlineID}:][]{FFlineID} which also presents FF results for the off-axis spectra within sparse observations, and a [CI] detection and deblending paper \citep[\citetalias{FFncc}:][]{FFncc}. 

The Feature Finder procedure is presented in \S\,\ref{sec:FF}, with details on the \herschel/SPIRE data products and observing modes given in \S\,\ref{sec:input}. The main FF steps are provided in \S\,\ref{sec:algorithm} including deriving the initial continuum estimate (\S\,\ref{sec:cont}), the iterative feature finding (\S\,\ref{sec:featureFinding}), incorporating frequency masking (\S\,\ref{sec:frequencyMasking}), limiting fitted line drifts (\S\,\ref{sec:limits}), estimation of the final signal-to-noise ratio, and removing spurious detections (\S\,\ref{sec:snr}). Details on the goodness of fit criteria and data flagging protocols are presented in \S\,\ref{sec:goff}. The methods to estimate the radial velocities, including compilation of radial velocities from the literature, are presented in \S\,\ref{sec:redshift}. In \S\,\ref{sec:neutralC} we describe a deblending method which improves the final feature results for sources with $^{12}$CO(7-6) and [CI]\,$\mathrm{^3P_2 - ^3P_1}$ lines. We also perform visual inspection of the FF products, summarised in \S\,\ref{sec:visual}. The FF data products are presented in \S\,\ref{sec:products}, and  \S\,\ref{sec:FFvalid} presents the main FF validation and results, including completeness validation testing (\S\,\ref{sec:completeness}), comparison with an alternative method (\S\,\ref{sec:pythonPeaks}), accessing the FF products (\S\,\ref{sec:access}), and comparison with known results for SPIRE calibration sources (\S\,\ref{sec:calibrators}). Summary and conclusions are provided in \S\,\ref{sec:summary}.  

The standard and highly processed SPIRE FTS data products have flux density units of Jy for point-source calibrated data, and intensity units of W\,m$^{-2}$\,Hz$^{-1}$\,sr$^{-1}$ for extended-source calibrated data. All SPIRE FTS spectra are presented as a function of frequency, in GHz, in the kinematic local standard of rest (LSRk) reference frame. The mnemonic designations of the two central detectors for the SSW and SLW arrays are \texttt{SSWD4} and \texttt{SLWC3} (see \citet{spire_handbook}).
\vspace{-18pt}
\section{The SPIRE FTS Spectral Feature Finder}\label{sec:FF}
The Spectral Feature Finder is developed in the  \herschel\ Interactive Processing Environment 
\citep[\texttt{HIPE; }][]{Ott10}, version 15. As mentioned in \S\,\ref{sec:intro}, the FF routine was developed as a preliminary analysis tool to help identify \herschel\ data products for follow-up investigation.  As the FF needs to process a large number of SPIRE FTS observations, it was developed to run automatically, while having enough flexibility to produce reliable results for a wide range of input spectra. 

\begin{figure*}
\centering
\includegraphics[trim = 0mm 0mm 0mm 0mm, clip,height=0.9727\textheight]{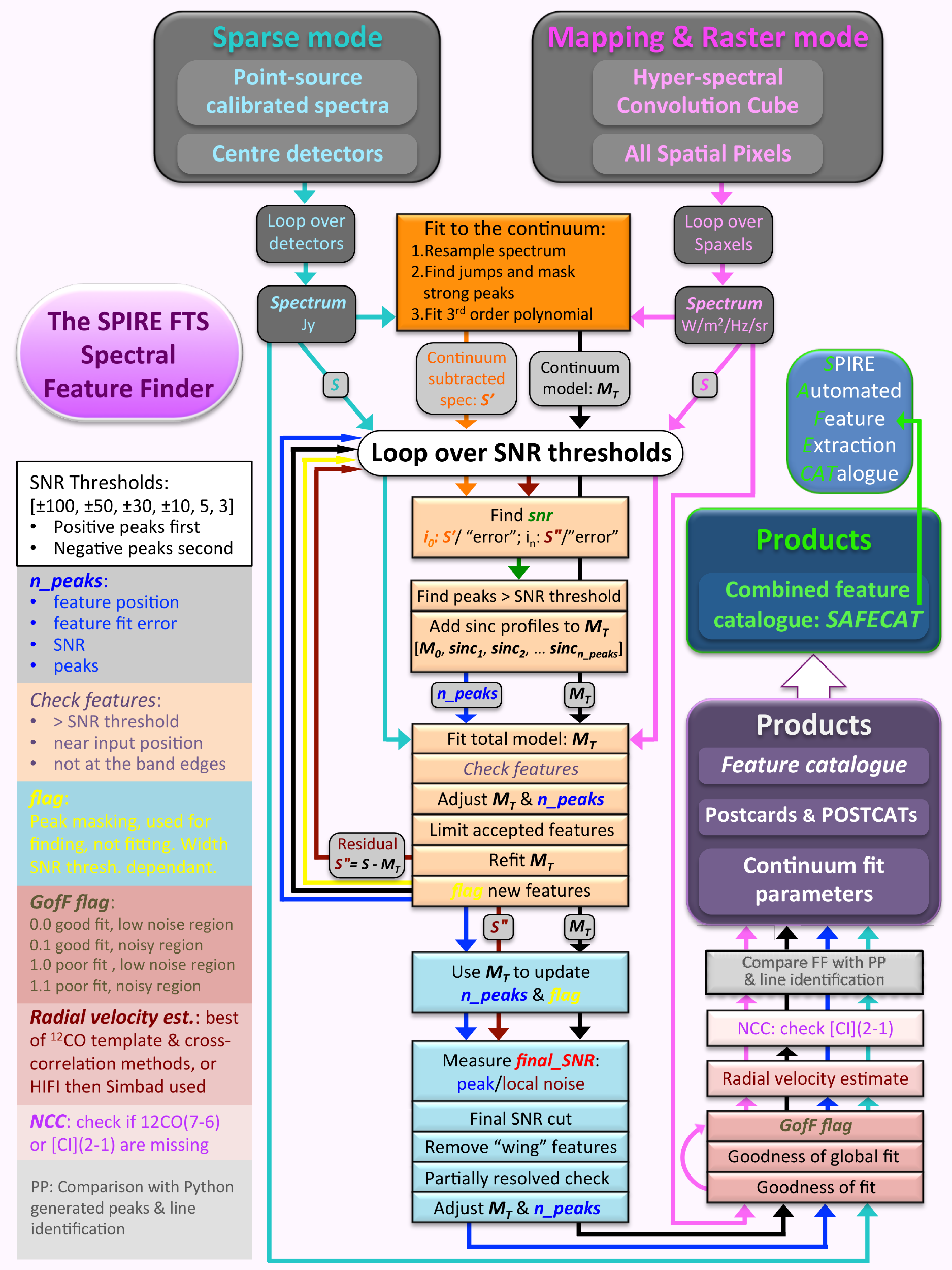}
\vspace{-6pt}\caption{The FF flowchart, which summarises the main FF steps. A detailed description is provided in \S\,\ref{sec:algorithm}.}
\label{fig:flowchart}
\end{figure*}

This section details each of the input products used, as well as the feature finder processing steps, the derivation of radial velocity estimates, and a dedicated check for in-band [CI] fine structure emission. The FF algorithm is depicted by the flowchart in Fig.\,\ref{fig:flowchart}.
\vspace{-12pt}
\subsection{Feature Finder input}
\label{sec:input}
From operational day (OD) 209 (08 Dec 2009) to the end of {\herschel}'s operational phase (OD~1447, 29 Apr 2013), there are 1802 SPIRE FTS observations available in the HSA, which total 2\,083.3\,hours. Before OD~209, SPIRE FTS data is not properly calibrated, as the final instrument settings were still being tested. The highest proportion of observations ($\sim$75\% by observation time) were taken in sparse, single-pointing mode (i.e., one SPIRE FTS footprint on-sky). Of those, 1162 were observed in HR mode (1\,455.6\,hours) and 330 in LR mode (70.2\,hours). There are 60 of these observations that were taken in a combined mode, where high and then low resolution scans are taken \citep[H+LR, 53.6\,hours, see][]{spire_handbook, Marchili:16,Marchili_HLR_MNRAS}. 
Tab.\,\ref{tab:HPDP} provides a breakdown of the SPIRE Spectrometer observations, including the number of observations, number of unique observing targets, and the hours of associated observing time. This is provided for several observation categories and classifications.
All the fully calibrated data products are presented in the HSA as spectral data sets, with one spectrum for each of the bolometer array detectors for sparse mode.

The SPIRE FTS also observed in intermediate (four jiggle positions), fully sampled (16 jiggle positions), and raster mapping modes, which make up $\sim$25\% of all observations at 558.4\,hours of integration time. There are two types of cube products available for mapping observations -- the Na\"{i}ve projection (NP) cube and Convolution projection (CP) cube.  The NP cube values are determined using the mean sum of the data falling within the spatial grid coordinates, and the CP cube values are determined using a weighted mean based on the proportional filling of the respective grid spaxel by the Gaussian beam profile of the associated detector element. See the SDRG for more details on NP and CP cubes and how they are created. Only CP cubes are considered by the feature finder, primarily because they tend to have higher SNR and do not contain empty spaxels in the middle of the maps as they are generated using more spatial information. There are, however, incomplete spectra at the edges of some CP cubes. These arise from the vignetting of the frequency dependent SPIRE beam (see \cite{Makiwa2013}) and result in frequency dependent spatial coverage. 

Fully calibrated mapping data are presented as hyper-spectral cubes, where the individual spectra have been projected onto equidistant rectangular RA/Dec grids. There are separate cubes for the long and short wavelength bolometer arrays with spaxels of different spatial extent, where each spaxel contains a single spectrum. The \citet[][]{spire_handbook} 
provides full details of SPIRE FTS observing modes and data products.

The primary input to the FF script is an FTS spectrum from either the SLW or SSW array. Standard SPIRE FTS data products, from the HSA, were used for most observations: the standard product generation (SPG) pipeline was run with \texttt{HIPE} version 14.1, and SPIRE calibration tree \texttt{spire\_cal\_14\_3}. The FF works on unapodized (i.e., unsmoothed) versions of the spectra, although for validating the results with an alternative detection method (see \S\,\ref{sec:pythonPeaks}) we also use apodized spectra. Both versions are available as standard products from the pipeline. While it is possible that some of the HPDP data employed in the FF processing falls under a different HIPE processing/calibration scheme, given that the FF outputs line frequency and SNR, and not integrated flux or intensity of spectral features, this was not observed to change the FF output.

The input data (and output products) depend only on observing mode and the spectral resolution of the associated observation. As introduced earlier, spectral resolutions include \textit{low} and \textit{high} spectral resolution (LR and HR, respectively) and  observing modes include \textit{sparse} and \textit{mapping} observations. Tab.\,\ref{tab:HPDP} provides an overview of the SPIRE Spectrometer observations available from the HSA and includes a breakdown of the number of observations in these different observing modes, as used within the FF. The table lists observations which were either restricted or otherwise excluded from the FF, as well as those which are included, and corresponding sub-categories. The restricted observations are special calibration observations, with non-standard detector settings, and are limited to 8 targets: Uranus, Mars, Neptune, OMC-1, DR\,21, IRC+10216, 4\,Vesta, and Dark\,Sky\footnote{Observations of an area of dark sky are set to match the longest science observation on a given observation day. The location of the SPIRE dark sky field is 17h40m12.00s, 69d00'00.00''.}.  Regarding the public observations excluded from the FF search, the Dark\,Sky observations are excluded, and there are 3 other featureless solar system objects (e.g., planets and asteroids) excluded from the FF search: Uranus, 1\,Ceres, and 4\,Vesta. The results for Neptune, Mars, and Saturn are also not included, as some of these lines are partially resolved and the FF, tuned specifically for unresolved spectral features, does not produce results as good as their existing models\footnote{\url{www.cosmos.esa.int/web/herschel/calibrator-models}}.
There is also one AFGL\,4106 observation excluded from the HR-sparse FF data due to a pointing error.  

Commentary on specific categories of SPIRE Spectrometer observation data as related to the FF is provided in \S\ref{sec:HRsparse}--\S\ref{sec:HPDPs}. 
\vspace{-12pt}
\subsubsection{High resolution sparse-mode observations}\label{sec:HRsparse}

For high resolution (HR) sparse-mode observations, the FF is configured to process the central detector of each array, as the large majority of these observations are of point-like sources or sources with little spatial extent. The FF processing of the off-axis detectors in sparse HR mode is presented in \citetalias{FFlineID}.
\vspace{-12pt}
\subsubsection{High resolution intermediate and fully sampled mapping observations}\label{sec:HRmap}

For HR intermediate and fully sampled mapping observations (and raster observations), the FF is run on every fully populated spaxel in both the SLW and SSW CP cubes. If the spectrum of any spaxel contains one or more instances of not-a-number (NaN) values, the FF skips that spectrum. 

\vspace{-12pt}
\subsubsection{Low Resolution Observations}\label{sec:LR}

The FF process is applied to low resolution (LR) observations, both for sparse and mapping mode. In this case, the iterative SNR loop is skipped and only the continuum parameters are derived for LR observing modes (see Tab.\,\ref{tab:mask}).
\vspace{-12pt}
\subsubsection{High+Low Resolution Observations}\label{sec:HRLR}

The H+LR observing mode is simply a sequence of HR and LR spectral scans. For the FF run, we treat each part independently and process it as either HR or LR mode, respectively. In Tab.\,\ref{tab:HPDP} these observations are included in the corresponding HR section to avoid double-counting with respect to observation totals.  Within the FF they appear in both the HR and LR sections.
\begin{table}
\begingroup
\newdimen\tblskip \tblskip=5pt
\caption{\label{tab:HPDP}SPIRE FTS observation classification and distribution. Results are provided for all SPIRE Spectrometer observations, and then split into the four main FF categories: Sparse HR, Mapping HR, Sparse LR, and Mapping LR. Indentations of the left-most column, and partial horizontal lines separating the numerical groups indicate subsets of the sub-category above the grouping. There are two observations that had a double pointing that were treated in the FF as double HR-sparse observations rather than single mapping observations; this is why the HR-sparse observations included in the FF are shown as 864 while the sub-groupings below this entry add up to 866. See \citetalias{FFlineID} for more details on the off-axis results.}
\nointerlineskip
\small
%
\newdimen\digitwidth
\setbox0=\hbox{\rm 0}
\digitwidth=\wd0
\catcode`*=\active
\def*{\kern\digitwidth}
\newdimen\signwidth
\setbox0=\hbox{+}
\signwidth=\wd 0
\catcode`!=\active
\def!{\kern\signwidth}
%
\tabskip=2em plus 2em minus 2em
\halign to \hsize{#\hfil& \hfil#\hfil& \hfil#\hfil& \hfil#\hfil\cr
\noalign{\doubleline}
Observation & Number & Unique & Obs.\cr
classification& of obs.& targets& time [h]\cr
\noalign{\vskip 2.5pt}
\noalign{\hrule}
\noalign{\vskip 5pt}

\textbf{Total}& 1\,802& 1\,088& 2\,083.3\cr
\noalign{\vskip -8pt}
 &\multispan3\hrulefill\cr
\hspace{12pt} Restricted& *\,*41& *\,**8& *\,*15.8\cr
\hspace{12pt} Public&     1\,761& 1\,088& 2\,067.5\cr
\noalign{\vskip -8pt}
 &\multispan3\hrulefill\cr
\hspace{24pt} Excluded from FF & *\,361& *\,**8& *\,329.8\cr
\hspace{24pt} Included in FF & 1\,400& 1\,084& 1\,737.7\cr  
\noalign{\vskip 2.5pt}
\noalign{\hrule}
\noalign{\vskip 5pt}

\textbf{Sparse High Resolution}& 1\,162& *\,681& 1\,455.6\cr
\noalign{\vskip -8pt}
 &\multispan3\hrulefill\cr
\hspace{12pt}Restricted& *\,*34& *\,**8& *\,*13.0\cr
\hspace{12pt}Public&     1\,128& *\,681& 1\,442.6\cr
\noalign{\vskip -8pt}
 &\multispan3\hrulefill\cr
\hspace{24pt}Excluded from FF& *\,260& *\,**8& *\,280.0\cr
\hspace{24pt}Included in FF&   *\,868& *\,676& 1\,162.6\cr 
\noalign{\vskip -8pt}
 &\multispan3\hrulefill\cr
\hspace{36pt}Continuum only& *\,199& *\,141& *\,439.7\cr 
\hspace{36pt}Point source&   *\,285& *\,169& *\,376.8\cr 
\hspace{36pt}Semi-Extended&  *\,258& *\,248& *\,266.2\cr 
\hspace{36pt}Extended source&*\,126& *\,126& *\,*79.9\cr 
\noalign{\vskip -8pt}
 &\multispan3\hrulefill\cr
\hspace{36pt}HPDP correction& *\,114& *\,*15& *\,*29.9\cr 
\hspace{36pt}Background sub.& *\,*33& *\,*32& *\,104.3\cr 
\hspace{36pt}H+L res.&        *\,*29& *\,*28& *\,*12.3\cr 
\hspace{36pt}off-axis&        *\,509& *\,290& *\,645.2\cr 
\noalign{\vskip 2.5pt}
\noalign{\hrule}
\noalign{\vskip 5pt}

\textbf{Mapping High Resolution}& *\,223& *\,202& *\,508.0\cr 
\noalign{\vskip -8pt}
 &\multispan3\hrulefill\cr
\hspace{12pt}Restricted& *\,**2& *\,**2& *\,**1.1\cr
\hspace{12pt}Public&     *\,221& *\,201& *\,506.8\cr 
\noalign{\vskip -8pt}
 &\multispan3\hrulefill\cr
\hspace{24pt}Excluded from FF& *\,*16& *\,**4& *\,*32.2\cr 
\hspace{24pt}Included in FF&   *\,205& *\,197& *\,474.6\cr 
\noalign{\vskip -8pt}
 &\multispan3\hrulefill\cr
\hspace{36pt}Raster&          *\,*15& *\,*14& *\,*57.1\cr
\hspace{36pt}Single Pointing& *\,190& *\,183& *\,417.5\cr
\noalign{\vskip -8pt}
 &\multispan3\hrulefill\cr
\hspace{48pt}H+L res.& *\,*31& *\,*31& *\,*41.3\cr 
\noalign{\vskip 2.5pt}
\noalign{\hrule}
\noalign{\vskip 5pt}

\textbf{Sparse Low Resolution}& *\,330& *\,187& *\,*70.2\cr
\noalign{\vskip -8pt}
 &\multispan3\hrulefill\cr
\hspace{12pt}Restricted& *\,**1& *\,**1& *\,**0.1\cr
\hspace{12pt}Public&     *\,329& *\,187& *\,*70.1\cr
\noalign{\vskip -8pt}
 &\multispan3\hrulefill\cr
\hspace{24pt}Excluded from FF& *\,*72& *\,**2& *\,*12.1\cr
\hspace{24pt}Included in FF&   *\,257& *\,185& *\,*58.0\cr
\noalign{\vskip 2.5pt}
\noalign{\hrule}
\noalign{\vskip 5pt}

\textbf{Mapping Low Resolution}& *\,*89& *\,*63& *\,*50.4\cr
\noalign{\vskip -8pt}
 &\multispan3\hrulefill\cr
\hspace{12pt}Restricted& *\,**4& *\,**3& *\,**1.5\cr
\hspace{12pt}Public&     *\,*85& *\,*63& *\,*48.8\cr
\noalign{\vskip -8pt}
 &\multispan3\hrulefill\cr
\hspace{24pt}Excluded from FF& *\,*15& *\,**2& *\,**6.3\cr
\hspace{24pt}Included in FF&   *\,*70& *\,*61& *\,*42.5\cr
\noalign{\vskip -8pt}
 &\multispan3\hrulefill\cr
\hspace{36pt}Raster&          *\,**5& *\,**2& *\,**4.3\cr
\hspace{36pt}Single Pointing& *\,*65& *\,*59& *\,*38.3\cr
\noalign{\vskip 2.5pt}
\noalign{\hrule}
\noalign{\vskip 5pt}
}
\endgroup
\vspace{-12pt}
\end{table}
\vspace{-12pt}
\subsubsection{Extended and semi extended sources observed in HR sparse-mode}\label{sec:semiEx}

Over the course of the \herschel\ mission, a number of sources with significant spatial extent were observed in sparse observing mode. For these observations, the FF was run with both the point-source and the extended-source calibrated processing pipelines. Due to the frequency dependent beam diameter of the SPIRE FTS \citep{Makiwa2013}, few sources are truly extended over the SPIRE bands and it is not always clear which of these calibration schemes will provide the better FF result. 
The number of HR sparse observations where the extended source calibration leads to better FF results is given in Tab.\,\ref{tab:HPDP}, as ``Extended source'' classification.

For some sources, neither the point nor the extended source calibration provide good spectral shape without a discontinuity in the overlap region or unphysical curvatures. In such cases, SECT \citep[][]{Wu2013} can be used iteratively to optimise the source size and shape in order to minimise the transition between the SLW and SSW data. This improves the spectral shape and continuum significantly, and consequently the feature detection and characterisation. For some calibration observations, HPDPs with semi-extended corrections exist (\S\,\ref{sec:HPDPs}) and were used instead of the standard HSA spectra. Therefore, a visual inspection (\S\,\ref{sec:visual}) was carried out to determine which calibration scheme was optimal for a given observation. 

Tab.\,\ref{tab:HPDP} lists the number of spectra visually classified as ``Semi-Extended'', based on the appearance of the SLW and SSW spectra. The input spectra were not corrected for the source shape and size with SECT, as this was considered outside the scope of the automated feature finder. The class is nevertheless provided to inform or warn the user that the FF results may be sub-optimal as neither the point nor the extended source calibrations are applicable.
\vspace{-12pt}
\subsubsection{The use of Highly Processed Data Products}\label{sec:HPDPs}
For some observations, including those observing calibration sources, there are expert produced data products, which are alternative or better than the nominal SPG output. Some of those are available as Highly Processed Data Products (HPDP\footnote{HPDPs are available in the \herschel\ Science Centre: \\ \url{https://www.cosmos.esa.int/web/herschel/highly-processed-data-products}}). For the FF input we use the following improved products:
\begin{enumerate}[leftmargin=*, label=\arabic*.] 
    \item A pointing offset correction as described by \citet[][]{Valtchanov2014}. This correction was only applied, as needed, to a subset of the calibration targets (only for HR sparse mode).
	\item A correction for partially extended sources using the Semi Extended Correction tool (SECT) in HIPE \citep[][]{Wu2013} (only for HR sparse mode).
	\item Subtraction of high background emission using averaged spectra from off-axis detectors within the same observation (only for HR sparse mode).\footnote{\herschel\ SPIRE Background Subtracted Spectra DOI: \href{https://doi.org/10.5270/esa-vrq3lz9}{doi.org/10.5270/esa-vrq3lz9}} 
	\item Spectral cubes undergo regridding that excludes the outer ring of vignetted detectors (typically for maps of faint sources, only for HR mapping mode).\footnote{\herschel\ SPIRE Re-Gridded Spectral Cubes DOI: \href{https://doi.org/10.5270/esa-77n5rc0}{doi.org/10.5270/esa-77n5rc0}}
\end{enumerate}	

For each SPIRE spectrometer observation operating in sparse mode, a separate set of background subtracted spectra (BGS
) exist, which are obtained by using the HIPE Background Subtraction tool in interactive mode. These HPDPs are used in place of the standard HSA products when FF results are improved, e.g., for
faint point-sources embedded in high Galactic cirrus. Meta data in FF products contain the parameters ``HPD USED'' when a pointing offset correction has been applied, and ``BGS USED'' when the alternate background subtraction has been employed.
The number of observations with these corrections is listed in Tab.\,\ref{tab:HPDP} as ``HPDP correction'' and ``Background subtracted''. 
%
%
%
%
\vspace{-12pt}
\subsection{Feature Finder algorithm}
\label{sec:algorithm}
The FF algorithm is depicted by the flowchart in Fig.\,\ref{fig:flowchart} and can be divided into four main components: 1) initial fit to the continuum, 2) iterative feature finding, 3) final SNR estimate and reliability checks, and 4) generating parameters for the features found. Each of these components is described in detail within this section.
\vspace{-12pt}
\subsubsection{Initial fit to the continuum}
\label{sec:cont}
Before starting to search for spectral features in the extracted spectrum, the FF fits a polynomial to the continuum in each of the two spectral bands: SSW and SLW. In order to ensure a good result, all significant features present in the spectrum need to be masked. To that end, the input HR spectrum is resampled onto a coarser, 5\,GHz, frequency grid (LR spectra are not resampled), and a ``difference spectrum'' is generated by subtracting adjacent flux values of the resampled spectrum. Discrete jumps in the difference spectrum, greater than 3.5 times the RMS, are interpreted as strong peaks in the unmodified input spectrum. A mask for the original spectrum is created, using a width of 30\,GHz centred on each of the strong peaks that have been identified. The original masked spectrum is then fitted using a 3$^{\rm rd}$ order polynomial for HR data and a 2$^{\rm nd}$ order polynomial for LR data. The reduced polynomial order for the LR case was selected to prevent the potential influence of sinc sidelobes on the continuum fit ($\sim$1\,GHz resolution for HR cf. $\sim$25\,GHz resolution for LR). The resulting best fit model of the continuum is used as a base for the spectral feature finding. In each iteration, newly identified features are added to a total spectral model including the continuum, and during the fitting procedure, the continuum parameters are allowed to vary. Throughout the FF process, fitting is always performed on the original spectrum.
\vspace{-12pt}
\subsubsection{Finding and fitting spectral features}
\label{sec:featureFinding}
The main finding and fitting of spectral features starts with the modelled continuum and the input spectrum. Each step in the iterative process applied is described as follows:
\begin{enumerate}[leftmargin=*, label=\arabic*.]
	\item The SNR spectrum is calculated by dividing the continuum subtracted spectrum by the associated {\it error} spectrum. The error spectrum is calculated by the SPIRE FTS pipeline \citep{fulton2016} as the weighted standard error of the mean signal from all spectral scans in an observation. 
	\item Regions with SNR greater than the iteration threshold are located and peaks are determined by merging all data points above the threshold within a 10\,GHz width window per peak.
	\item For each peak, a sinc function profile (see Fig.\,\ref{fig:flagWidth} and the SDRG) is added to the total model (which starts before the first iteration with the polynomial fitted to the continuum, and builds as the process iterates). The width parameter of the sinc function is kept fixed to the unresolved HR instrumental line width of \mbox{$1.2/\pi = 0.382$ GHz}. This is because the majority of spectral features in SPIRE FTS HR data are unresolved. Partially resolved sources should be fitted using a sinc convolved with a Gaussian, however, allowing the width to be a free parameter does not generally provide a better fit to a partially resolved line (in practice, trials during the FF development cycle demonstrate that the width hardly changes).
	\item A global fit is performed.  The features found are assessed against the following set of metrics and discarded from the total model if:
		\begin{enumerate} 
			\item the line amplitude parameter has the incorrect sign, 
			\item the SNR is $<$ -500 (for the negative sweep only),
			\item the fitted position has moved more than 2 GHz from the line's starting position,
			\item the $|$SNR$|$ is $<$ $|$SNR threshold$|$, or
			\item the feature is within the fringe avoidance regions (10 GHz at the end of the bands).
		\end{enumerate}	
	\item A check is then performed to reduce the number of multi-sinc fits to partially resolved lines.
		\begin{enumerate}
			\item The position of each new feature is compared to features already identified in previous iterations of the algorithm, under the likely assumption the previous features have higher SNRs.
			\item If within 1.2 GHz of any neighbouring feature, the new feature is rejected and the position of the previous feature is set back to its position within the previous global fit.
			\item The position of the existing stronger feature is assumed to be at the peak of a partially resolved or highly blended feature or a feature whose width has been significantly affected by noise.
		\end{enumerate}
	\item Surviving features are assumed to be valid with a stable fitted position, so a positional restriction of 2 GHz (i.e., 5-$\sigma$) is applied to their corresponding sinc model for subsequent iterations.
	\item At this point the global fit is repeated.
		\begin{enumerate}
			\item If the fit succeeds, new features are added to the list of detected features and a mask is applied to the spectrum on either side of the detected feature with a frequency width dependant on the SNR threshold (see \S\,\ref{sec:frequencyMasking}). No new features are accepted if they fall within a masked frequency region.
			\item If the fit fails, all new features are removed and the results remain unchanged for input into the next iteration at a lower SNR threshold.
		\end{enumerate}
	\item A residual is calculated by subtracting the best-fit total model (including both the identified spectral features and the continuum) from the original spectrum, and it is passed to the next iteration of feature finding. The residual is used for the SNR calculation when searching for new peaks.
	\item Once the loop has cycled through the strongest emission features down to the lowest SNR threshold, the loop is repeated searching for absorption features, again starting with the strongest and working down to the final SNR threshold (by definition, SNRs are negative for absorption features, so the threshold evaluation is performed on the absolute SNR magnitude).
\end{enumerate}
\vspace{-12pt}
\subsubsection{Frequency masking}
\label{sec:frequencyMasking}
%
%
In order to lessen the number of spurious detections that can occur in the wings of another feature, a spectral frequency mask is created and used during the peak detection and fitting process. The mask is built upon and updated after each iteration and then carried forward and used in subsequent iterations. 
%

\begin{figure*}
\centering
\makebox[\textwidth][l]{
\includegraphics[trim = 0mm 0mm 0mm 0mm, clip,width=0.5\textwidth]{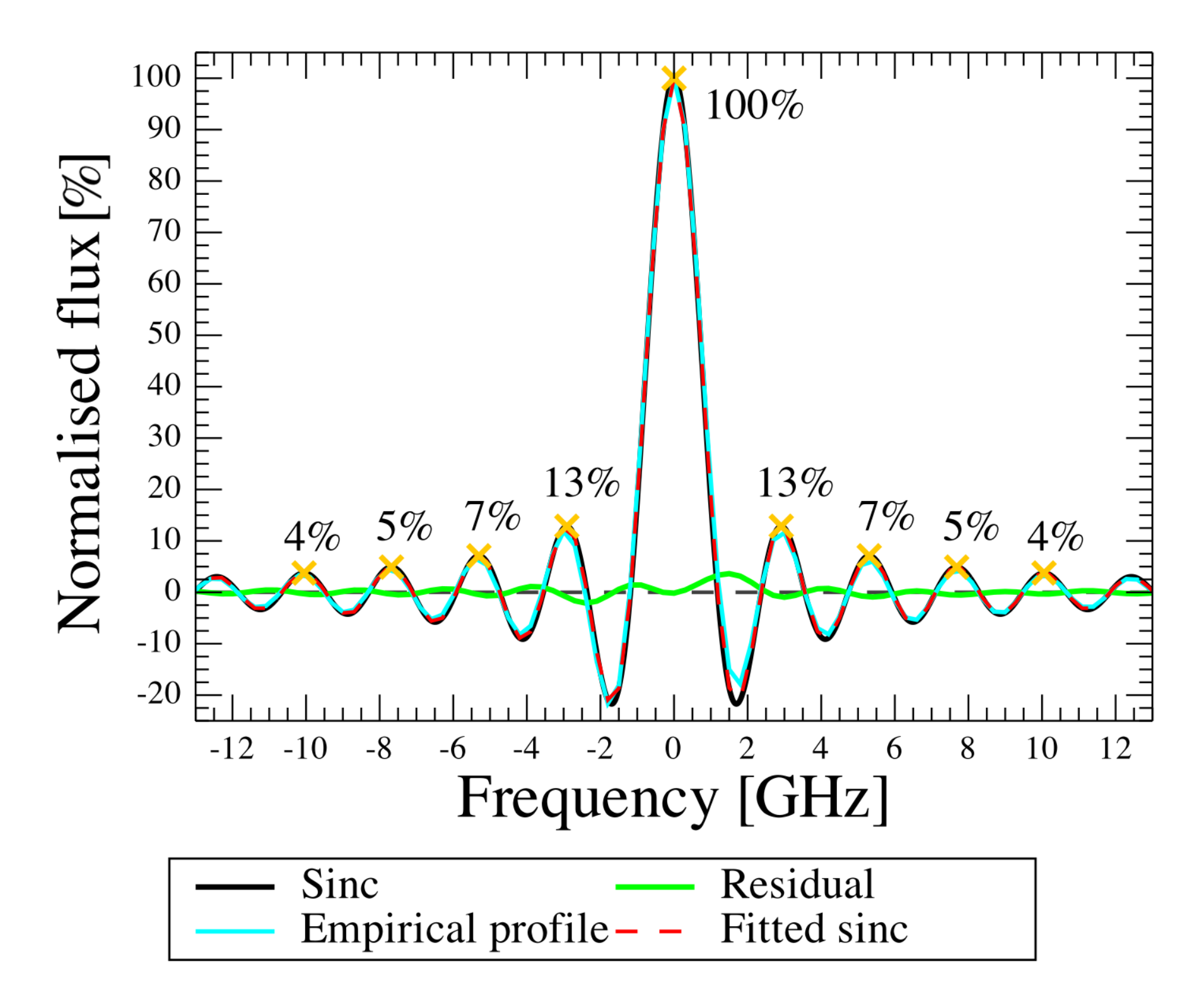}
\includegraphics[trim = 0mm 0mm 0mm 0mm, clip,width=0.5\textwidth]{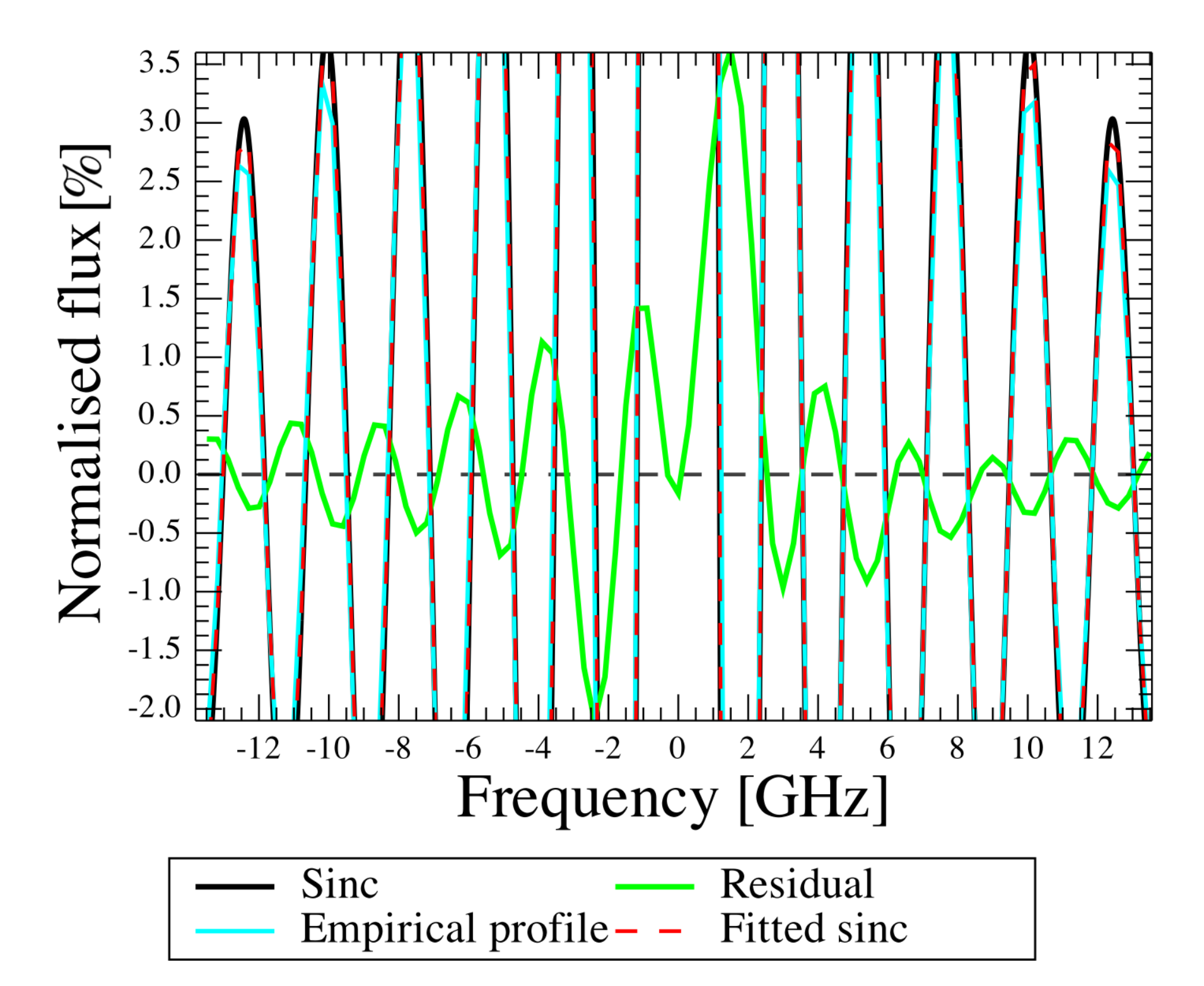}
}
\vspace{-18pt}
\caption{An ideal feature (a pure sinc function; black line) with a SNR of 100 with the relative strengths of the positive side-lobes indicated. Also plotted are the empirical line profile (cyan line, see \citetalias{Hopwood15}) and the residual (green line) after fitting and subtracting a sinc (red dashed line) from the empirical line profile. The right plot is a zoomed in version of the left plot, to more clearly show the low level of the residual, despite the line asymmetry present.}
\label{fig:flagWidth}
\vspace{-12pt}
\end{figure*}

Each masked region is centred on a detected feature, with the masking width dependent on the iteration SNR threshold. The widths for each threshold were determined by the expected side-lobe amplitudes of a pure sinc function. For example, Fig.\,\ref{fig:flagWidth} shows the relative strength of the side-lobes for an ideal SPIRE FTS instrument line shape (ILS) emission line, with SNR of 100 and a flat associated error of 1. 
From this example, for features with SNR\,$>$\,100, the 3$^{\rm rd}$ positive side-lobes will appear to have a SNR\,$>$\,5 corresponding to a frequency difference from the central peak of $\sim$8\,GHz.  Fig.\,\ref{fig:flagWidth} also shows the empirical SPIRE FTS ILS, which was obtained by stacking CO features from calibration observations of NGC\,7027 \citepalias{Hopwood15}. 

In setting a mask-width, some additional considerations are necessary, as in reality the noise spectrum will not be flat and the fit to the feature will not be perfect. Strong and unresolved features, significantly above the random noise, tend to be well fitted. Even a small misalignment of the true feature frequency and the model best-fit (e.g., due to the mild asymmetry of the observed empirical SPIRE FTS ILS, \citealt[][]{Naylor:16}), or widening of the peak (due to velocity dispersion of the source or noise) may leave prominent spectral residuals in the wings. 

The iterative mask widths used, and their corresponding SNR thresholds are provided in Tab.\,\ref{tab:mask}. The final two widths (both 2\,GHz) may appear overly conservative, however, this is the necessary minimum, due  to other constraints in the FF flow (see \S\,\ref{sec:limits}). Here we have assumed that the new fitted peak amplitudes are not stronger than the previous iteration SNR threshold and that the SNR does not significantly change when the final measurement (and SNR) is calculated (see \S\,\ref{sec:snr}), which is generally true for the strongest features present. To improve the reliability of the FF products by removing false detections, the negative $-5$ and $-3$ SNR thresholds are omitted for the absorption feature search.
%
%
\begin{table}
\begingroup
\begin{center}
\newdimen\tblskip \tblskip=5pt
\caption{\label{tab:mask}Iterative FF SNR thresholds.}
\nointerlineskip
\small
%
\newdimen\digitwidth
\setbox0=\hbox{\rm 0}
\digitwidth=\wd0
\catcode`*=\active
\def*{\kern\digitwidth}
\newdimen\signwidth
\setbox0=\hbox{+}
\signwidth=\wd 0
\catcode`!=\active
\def!{\kern\signwidth}
%
\tabskip=2em plus 2em minus 2em
\halign to \hsize{\hfil#&\hfil*#\hfil& \hfil#\hfil& \hfil#*\hfil&#\hfil\cr
 &\multispan3\hrulefill& \cr
\noalign{\vspace{-8.0pt}}
 &\multispan3\hrulefill& \cr
& Iteration &SNR threshold &Mask width [GHz]& \cr
\noalign{\vspace{-5.5pt}}
 &\multispan3\hrulefill& \cr
&1& $\pm$\,100& 8& \cr
&2& $\pm$\,*50& 8& \cr
&3& $\pm$\,*30& 5& \cr
&4& $\pm$\,*10& 4& \cr
&5&     !\,**5& 2& \cr
&6&     !\,**3& 2& \cr
\noalign{\vspace{-7.5pt}}
 &\multispan3\hrulefill& \cr
\noalign{\vspace{-12.5pt}}
}
\end{center}
\endgroup
\vspace{-12pt}
\end{table}
The right panel of Fig.\,\ref{fig:flagWidth} is a vertically zoomed in view of the left panel which shows more clearly the residual after fitting a sinc to the empirical line profile.  This plot indicates that for a strong and unresolved feature, the residual has little influence at SNRs\,$>$\,5. However, that conclusion does not take into account blending, strong noise features, or partially resolved lines.
\vspace{-12pt}
\subsubsection{Limiting spectral feature drift}
\label{sec:limits}
The frequency mask for identified features is carried forward to the next iteration.  Newly identified features added to the working model are not allowed to have frequencies within the masked regions.  This masking, however, does not prevent features from wandering into the forbidden regions during the parametric minimisation process within the global fit portion of the FF algorithm. Through empirical study of the minimisation routine, the fitting engine \citep[][]{fitter} does not respond well with strict frequency limits imposed upon all new features, with the fit much less likely to converge in such cases. This means that during the global fitting procedure, the fitted position of newly added sinc profiles may drift into the forbidden zone and fit an unwanted side-lobe or noise feature. Therefore, to minimise such movement and reduce the number of spurious features, a number of checks are made after the first global fit for new features found. These checks are listed in \S\,\ref{sec:snr}
. Any surviving new feature is assumed to be stable enough for its position to be restricted to $\pm$\,2\,GHz about its fitted frequency. A final check for features that have drifted to fit a side-lobe of another feature, or if the feature is a partially resolved line, is made after the final global fit. These final checks mean that features may be removed without refitting the total model as their removal has little impact on the final positions and SNRs of the main and neighbouring features.
\vspace{-12pt}
\subsubsection{Removal of spurious sidelobe features and estimation of the final SNR}
\label{sec:snr}
The residual in Fig.\,\ref{fig:flagWidth} was used as a rough guide to set the search region when checking for spurious detections in the wings of strong features, which is carried out after the main loop has completed. Thus, a validation step is performed to characterise each identified feature candidate as either a valid spectral feature, or a false fitting of a sidelobe to a prominent spectral feature. 
\begin{enumerate}[leftmargin=*, label=\arabic*.]
	\item The SNR is calculated as $A_{\mbox{fit}}/\sigma_{A}$, where $A_{\mbox{fit}}$ is the best-fit amplitude of the spectral feature, and $\sigma_{A}$ is the standard deviation of the residual spectrum, determined for a spectral region in the neighborhood of the feature and following localized baseline subtraction. 
	\item Although the final SNR threshold is set to 5, the iterative line searching is actually conducted down to a SNR threshold of 3 in the main loop, with the application of an SNR cut off at  5, following the final step of the iterative search. This approach is used to avoid missed features and false positives as the final SNRs for each feature are subject to change during the iterative line search. 
	\item Identified features (i.e., with $|$SNR$|$ $\ge$ 5) are checked for proximity to other features to see if they are likely fitting a sidelobe or peak in the wings of a more prominent feature. The final consideration when checking for wing-fitting, is the SNR of the identified main peak compared to that of the potential spurious feature. Considering again the ideal line shown in Fig.\,\ref{fig:flagWidth}, the main peak would need to have an SNR greater than 23 for the 1$^{\rm st}$ positive side-lobe to have an SNR above the final SNR search threshold. However, leaving room for blending, partially resolved lines, and noise (and remembering peak positions can move during the global fit), limiting the minimum SNR of the main peak to 10 leads to few sidelobe-fit features falsely identified as spectral features. The final condition before a feature is labelled as false is the absolute ratio of the main peak to wing SNRs, which is required to be greater than four for the feature to be rejected. 
	\item Due to the mild asymmetry in the observed SPIRE FTS ILS \citep[][]{Naylor:16}, the search window is asymmetrically centred about the feature location, starting 2\,GHz below and ending 2.5\,GHz above the central frequency for all SSW spectral features.  
	\item The flagging regions in the SLW array are modified to improve the positive detection of [CI]\,$\mathrm{^3P_2 - ^3P_1}$ fine-structure emission near the $^{12}$CO(7-6) rotational transition feature, against false features fit to sinc sidelobes (rest frame $\sim$806.65\,GHz, $\sim$809.34\,GHz, for $^{12}$CO and [CI], respectively, see \S\,\ref{sec:neutralC}, \citetalias{FFncc}, and \citealt{Scott16NCC}). Thus, the SLW search window starts between 1.5--2\,GHz below (1.5\,GHz for the $^{12}$CO(7-6) feature, 2.0\,GHz otherwise), and ends 2.5\,GHz above the central frequency. This reduced SLW range results in a high success rate of detecting the neutral carbon emission of interest, without leaving spurious wing-fit features. To further improve detection, a dedicated search for the neutral carbon features is performed after the main loop is complete (see \S\,\ref{sec:neutralC} and \citetalias{FFncc}).
        \begin{enumerate}
			\item 
			For each positive feature there is a search at lower frequencies of the prominent peak. If a more prominent peak is found with SNR $>$ 10, and the SNR$_{\mbox{strong}}$/SNR$_{\mbox{weak}}$ ratio is $>$ 4 then the weaker peak is labelled as a wing fit and discarded. For a discarded feature, no adjustment is made to the nearby existing features, on the assumption that the discarded wing peak, if any, is significantly weaker than the main peak. In the SPIRE FTS data, typically the [CI] feature is of weaker intensity than the neighbouring CO feature, hence the asymmetric window function for this region of the SLW band. 
			\item The search for a stronger peak is then repeated to the higher frequency side, this time using the absolute SNR, so negative features are also considered. 
			The same SNR checks are performed as in the previous step.
		\end{enumerate}
\end{enumerate}
After the main loop has completed, the final residual is used to determine the final SNR for each identified feature.
\vspace{-12pt}
\subsection{Goodness of fit and the feature flags}
\label{sec:goff}
To assess the confidence level of an FF-identified spectral feature, a quality metric has been developed that combines the goodness of fit for each feature found (GoF) with the Bayesian ratio B10 \citep{bernardo1994bayesian} of the total model. GoF and B10 are used in conjunction with frequency regions identified as suffering higher than average noise to assign a flag to each feature (FF flag). 
\vspace{-12pt}
\subsubsection{Goodness of Fit}
\label{sec:gof}
The $\chi^2$ statistics is commonly used as a goodness of fit metric. It is, however, only appropriate when the associated noise of the data considered is Gaussian in nature. SPIRE FTS data have a high incidence of systematic noise, such as fringing due to residual telescope and instrument emission, and ringing due to the spectral features themselves (i.e. the sinc wings). The ringing is stronger for stronger lines, which leads to $\chi^2$ tending to higher values for higher SNR lines. $\chi^2$ is also sensitive to the continuum. The feature finder requires a goodness of fit metric that is not sensitive to the continuum or line flux. Such a metric, {\it r}, can be calculated using a cross-correlation function between fitted feature and total model, as\\
\vspace{-12pt}
\begin{equation}
\label{eq:r}
r  = \frac{1}{N} \sum_{|\nu-\nu_0|<R}\frac{[S(\nu) - \bar{S}_R][M(\nu) - \bar{M}_R]}{\sigma_{S}\sigma_{M}},
\end{equation}
where {\it R} is the frequency range for the estimation, $\nu_0$ is the line frequency, {\it N} is the number of frequency bins within the frequency range, $S(\nu)$ is the flux density at frequency $\nu$, $M(\nu)$ is the model at frequency $\nu$, $\bar{S}_R$ and $\bar{M}_R$ are the average flux and model within the frequency range, with $\sigma_{S}$ and $\sigma_{M}$ the corresponding standard deviations within the frequency range. Based on testing the FF with spectral simulations and spectra of SPIRE calibration targets, a fit is taken as good for $r > 0.64$ and poor for $r < 0.64$.

The variable {\it r} is not sensitive to the continuum, due to the subtraction of local averages. Furthermore, $r$ is not sensitive to line flux, because of division by the standard deviations. Comparing the GoF to the SNR of features found for 150 calibration observations, shows that although there is not a perfect correlation, there is a strong trend of higher {\it r} for higher SNR features. Regardless of SNR, {\it r} saturates at $\sim$\,0.95, which may be due to line asymmetry. Strong features that have low {\it r} may be partially resolved and/or suffer from close strong features, i.e. high levels of sinc-wing blending. Similarly, {\it r} is not low for all weaker features. These features may be in low noise areas and be relatively isolated, i.e. low levels of sinc-wing/sidelobe blending.
\vspace{-12pt}
\subsubsection{Bayesian ratio B10}\label{sec:tofe}
The second goodness of fit metric for the FF is the Bayesian ratio B10, which we define as follows:
\begin{equation}
\mathrm{B10} = \log{P(D|H_1)} - \log{P(D|H_0)},
\end{equation}
where $P(D|H_1)$ is the evidence for the total model $H_1$ with the putative sinc-function feature in and $P(D|H_0)$ is the evidence for a model without it. The logarithm of each model evidence is returned by the fitting engine, assuming flat priors on the model parameters and taking into account the noise scale (see \citealt{fitter} for details). If the two models are equally good at describing the data then B10 will be zero. In this case it is highly likely the feature is spurious.  As this is a probabilistic check, no assumption about the noise is made and therefore B10 is insensitive to systematic noise. An empirical threshold on B10 of -6 is set such that a B10 difference of -6 or greater indicates a true detection, while a difference of less than -6 is associated to a spurious detection. The Bayesian ratio test is complimentary to {\it r}.

Generally, most features are labelled as ``True'' by B10, if their SNR is greater than 15.  B10 tends to fail when assessing the model fitted to a spectrum that has many spectral features, so it is only applied when there are fewer than 35 features found. The algorithm implementation is also time consuming and is therefore turned off for mapping observations, where the FF has to evaluate {\it many} more spectra per observation.

Bearing in mind that B10 aims to distinguish between true and false detections, while GoF aims to evaluate the quality of the fit, the results of GoF and B10 were compared for the same 150 calibration observations previously mentioned. For the 1731 features found by the FF, 1259 (70\%) were deemed as good detections by both, with 46 (3\%) labelled as poor by both. There was disagreement between the two methods for 137 features, while 335 features fell into an ambiguous zone for both. A description of how these two metrics are used together in the FF is in \S\,\ref{sec:featureFlags} below.
\vspace{-12pt}
\subsubsection{Feature Finder flags}
\label{sec:featureFlags}
\begin{figure}
\centering
\includegraphics[trim = 0mm 0mm 0mm 0mm, clip,width=0.926825\columnwidth]{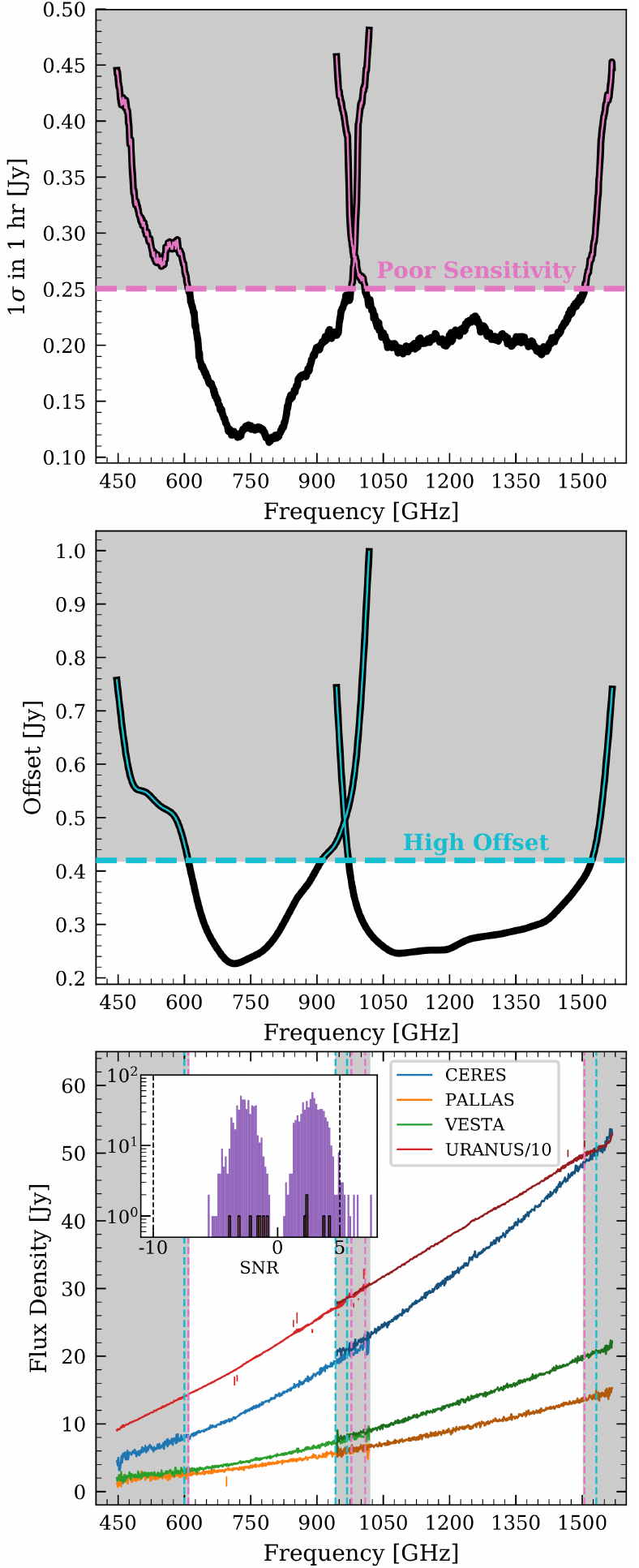}
\vspace{-6pt}
\caption{\label{fig:noisy}Noisy regions (shaded), defined using the SPIRE FTS point-source sensitivity and continuum offset (top, and centre, respectively, see \citetalias{Hopwood15}). 
The bottom panel shows the average spectra, and associated FF results, from SPIRE FTS calibration observations of each of the featureless sources Ceres, Pallas, Vesta, and Uranus (scaled by 10). Although no spectral features were identified above the SNR cutoff thresholds, the 12 low SNR features that were identified are indicated with vertical ticks (as is done in the nominal FF postcards, see Fig.\,\ref{fig:postcardSparse}) and in the inset histogram (bars with black outline). The inset histogram also shows the FF results on the \textit{individual} HPDPs for these sources (bars without outline), the distribution of which confirms the FF low SNR cutoff thresholds. 
}
\end{figure}

The GoF and B10 criteria are combined with ``noisy'' regions in the frequency range of both detector bands to define a FF flag. The SPIRE FTS sensitivity and 1\,$\sigma$ additive error on the continuum (known as the continuum offset) were used along with spurious detections in repeated SPIRE FTS calibration observations of featureless sources to define these ``noisy'' regions. Fig.\,\ref{fig:noisy} shows the point-source calibrated sensitivity and continuum offset taken from \citetalias{Hopwood15}, along with spurious features present in the FF results for Uranus, Ceres, Pallas and Vesta. There is significant noise seen at the low frequency end of SLW, due to imperfect instrument thermal background subtraction. Both the sensitivity and continuum offset steeply increase at the edges of the bands. These characteristics are similar for both types of noise and were considered along with the spurious features found for featureless sources, to determine the noisy regions as $< 608.6$\,GHz and $> 944.0$\,GHz for SLW and $< 1017.8$\,GHz and $> 1504.7$\,GHz for SSW.

There are four combined FF flags:
\vspace{-6pt}
\begin{itemize}
	\item 0.0: a good fit in a lower noise region
	\item 0.1: a good fit in a noisy region
	\item 1.0: a poor fit in a lower noise region
	\item 1.1: a poor fit in a noisy region,
\end{itemize}
\vspace{-6pt}
\noindent where the flag for goodness of fit is 0 (a good fit) or 1 (a poor fit). 0.1 is added to this flag if a feature is found within a noisy region.
A fit is deemed good when both GoF and B10 ``good'' criteria are met. If there are no B10 results, due to a large number of features found in a single spectrum, or for mapping observations, then GoF alone is used. %
\vspace{-12pt}
\subsection{Source radial velocity estimate}\label{sec:redshift} 
Here we briefly describe two different approaches able to provide a source radial velocity estimate for the FF catalogue entries, method 1 is based on the identification of the characteristic $^{12}$CO rotational ladder (rest frequencies for the lines within the CO ladder are provided in the line template at the top of Figs.\,\ref{fig:calibratorsHitMiss} and \ref{fig:speciesHitMiss}), and method 2 is based on cross-correlation of the FF catalogue with a spectroscopic line template catalogue based on HIFI data and included within HIPE. Both methods are detailed in full in the corresponding companion FF paper \citepalias{FFredshift}. 

Method 1 relies on the abundance of $^{12}$CO in the interstellar medium and the strong infrared ladder of its rotational transitions. There is also a near even spectral distribution of the $^{12}$CO emission, with a (rest-frame) difference between the spectral lines of $\sim$115.1\,GHz, and a variation less than the resolution of HR SPIRE FTS spectra (i.e., 1.2\,GHz). The list of features found after completion of the main loop is searched for the presence of the $^{12}$CO ladder. This is accomplished by searching an array of frequency differences between detected features for elements that are an integer multiple of $\sim$115.1\,GHz.  The set of features with the highest number of matches is taken as the most likely $^{12}$CO ladder candidate set. The chosen set of $^{12}$CO candidates are paired with their nearest rest frequency transition and are used to derive a velocity estimate. Candidate features producing high standard deviations in the per-candidate feature velocity estimates are removed. The number of identified $^{12}$CO features remaining is used to assess the quality estimate with more identifications producing a higher quality estimate.  When $^{12}$CO is not identified within a candidate spectrum, but there is \textsc{[NII]} emission (the $\mathrm{^3P_1 - ^3P_0}$ transition, rest frequency 1\,461.13\,GHz) 
with SNR\,$> 10$, this feature alone is used to estimate the velocity and the result given an appropriate quality flag.

Method 2 cross-correlates the list of features with a template catalogue (see e.g. \citealt{Zucker03}). The template catalogue contains most of the far-infrared characteristic lines for astronomical sources and is available as part of HIPE. The two catalogues are put on the same frequency grid and cross-correlated for a range of radial velocity offsets. The velocity for which the cross-correlation signal is maximised is taken as an initial estimate. Performing an initial line identification using the template catalogue further improves the radial velocity estimate. In some cases, due to degeneracies from the $^{12}$CO ladder or other molecular lines, the highest peak of the cross-correlation is not the actual radial velocity. That is why, out of the five strongest cross-correlation peaks, we select the one that maximises the number of identified lines.

In addition to the two methods, we also compiled radial velocity estimates from CDS/Simbad database and a list of radial velocities collected for the HIFI instrument (Beniamatti et al. 2017, private communication) and \citet[][]{wilson17}. The FF catalogue contains only one radial velocity estimate, its associated uncertainty, and a flagging parameter indicating how the estimate was derived. The FF methods take priority when the resulting radial velocity estimates return metrics indicating high quality. Estimates from external references are used when our methods fail or return low quality metrics.

Only method 1 is used when estimating radial velocities in mapping observations as this method is more rigorously validated and computationally efficient. To maintain consistency with sparse observation, detected feature lists from the SLW and SSW bands are combined in spatially overlapping regions. Since the SLW and SSW cubes express different on-sky footprints with spaxels of different spatial dimensions, feature lists are combined in the following way. An SSW equivalent grid is generated with each element populated by a feature list of the detected SSW features from the corresponding SSW spaxel spectrum. For each grid element, the SLW feature list is taken from the detected features of the spectrum in the nearest SLW spaxel. If the nearest SLW spaxel is more than $\sqrt{2}$ times the width of an SLW spaxel away from the grid element, then no SLW features are projected into the grid element. The resulting feature list for each grid element is input to method 1 with the end product being a velocity map. The peripheries of these maps tend to contain only SLW features due to the different cube footprints.
%
\vspace{-12pt}
\subsection{Neutral carbon check}\label{sec:neutralC}
Although the checks for wing-fitting are geared towards preserving the detection of [CI]\,$\mathrm{^3P_2 - ^3P_1}$ in the SLW spectra, it is not always positively identified within the primary FF algorithm.  As mentioned above, the [CI]\,$\mathrm{^3P_2 - ^3P_1}$ fine structure transition is in close spectral proximity to the $^{12}$CO(7-6) rotational transition and these features show significant line blending. As this is a prevalent feature in far-infrared spectra of certain source types, a dedicated search for missed [CI] is performed after the source radial velocity ($v$) is estimated. The method assumes that both [CI]\,$\mathrm{^3P_2 - ^3P_1}$ and $^{12}$CO(7-6) are coupled in approximately the same reference velocity frame, giving a separation between the two features of 2.69\,GHz (rest frame). For a non-zero radial velocity, $v$, this difference can be expressed as
\vspace{-3pt}
\begin{equation}
\label{eq:NCCdiff}
\Delta = \left(1 + \frac{v}{c} \right)^{-1}\,\times\,2.69\,\,\,[\mathrm{GHz}],
\end{equation}
where $c$ is the speed of light.

The $^{12}$CO(7-6) emission feature, although typically dominant in terms of intensity compared to the [CI]\,$\mathrm{^3P_2 - ^3P_1}$ emission, is not necessarily stronger than [CI]\,$\mathrm{^3P_2 - ^3P_1}$, therefore a search is performed that looks for features within $\pm0.8$\,GHz of the velocity corrected positions of the $^{12}$CO(7-6) and the [CI]\,$\mathrm{^3P_2 - ^3P_1}$ features. If only one emission feature is found by the nominal FF line identification tool, a section of the input spectrum of width 40\,GHz and centred on the frequency of the detected feature is extracted. The partial spectrum is then fitted for three test cases, using one or two sinc functions and the polynomial continuum model.  Each of the three residuals is then assessed for improvements with the best case selected as the most likely scenario. Firstly, a single sinc is fitted (assuming only one feature is present). Secondly, assuming the $^{12}$CO(7-6) is dominant and that [CI]\,$\mathrm{^3P_2 - ^3P_1}$ may not yet have been identified, two sincs are fitted, the first positioned at the primary feature and the second offset to higher frequency by an amount $\Delta$ and set to 30\% of the amplitude of the first. Thirdly, assuming that the [CI]\,$\mathrm{^3P_2 - ^3P_1}$ feature is dominant, two sincs are fitted again, with the secondary sinc offset to lower frequency by $\Delta$. The minimum root-mean-squared residual of the localized spectra dictates the best of the three scenarios. If the minimum residual result is either case 2 or 3, the estimated [CI]\,$\mathrm{^3P_2 - ^3P_1}$ or $^{12}$CO(7-6) line is added to the feature catalogue. 

If the routine determines that [CI]\,$\mathrm{^3P_2 - ^3P_1}$ is detected, either by the main loop or as a result of the procedure outlined above, a search for [CI]\,$\mathrm{^3P_1 - ^3P_0}$ (rest frequency 492.16\,GHz) initiates. Again a section of the input spectrum is modelled using a sinc function representing the [CI]\,$\mathrm{^3P_1 - ^3P_0}$ feature and the polynomial continuum model. The center of the spectral section and the initial estimate for the emission frequency of the model feature is derived from the detected frequency of the [CI]\,$\mathrm{^3P_2 - ^3P_1}$ feature. A successful detection occurs if the fitted feature returns a SNR $\ge$ 5, and the emission frequency is within 0.8\,GHz of the predicted frequency based on the [CI]\,$\mathrm{^3P_2 - ^3P_1}$ line. Further details about the neutral carbon check routine, and examples of each of the three test cases, are presented in \citetalias{FFncc}.
\vspace{-12pt}
\subsection{Visual inspection of the FF results}
\label{sec:visual}
In total, 868 HR sparse observations were processed by the FF. The spectra for 384 of these exhibit characteristics associated with sources that have some spatial extent (see the Semi-Extended and Extended source entries in Tab.\,\ref{tab:HPDP}). For these 384 observations, there can be a poor shape to the continuum and high levels of fringing in the point-source calibrated data, and thus the point-source FF results may not be optimal. The FF results for all 868 observations were visually inspected, with an additional inspection of the extended-source calibrated FF results for the 384 partially or fully extended sources that were observed in sparse mode. The results were graded as ``good'', ``acceptable'' or ``poor'', depending on the number of features found or missed and the polynomial fit to the continuum. Overall, and considering both the point and extended-source calibrated data, 99\% of the FF results fall into the ``good'' or ``acceptable'' categories. 8\% of these show an improved catalogue for the extended-source results, in one or both of the centre detectors. All partially or fully extended sources are flagged in the FF catalogue(s) meta data. If the alternate calibration results for one or more of the centre detectors improves beyond the point-source results, this is also indicated in the metadata.

\vspace{-18pt}
\section{The Feature Finder products}\label{sec:products}
The FF produces products for all publicly available SPIRE FTS observations, although some targets known to be featureless are skipped, this includes observations of Uranus, asteroids, Callisto, and the SPIRE dark sky field \citep[][]{Swinyard2014, Benielli2014, Muller13, SpencerDark:16}. The type of FF product does vary depending on the resolution and observing mode, for instance, there are few features for LR observations, so there are no feature catalogues provided for these, but the postcards and continuum parameters are provided in the FF data products. The results for Neptune, Mars, and Saturn are also not included, as some of these lines are partially resolved and the FF, tuned specifically for unresolved spectral features, does not produce results as good as their existing models. 

The FF products provided (per observation) are:
\vspace{-6pt}
\begin{itemize}
	\item a significant spectral feature catalogue (for HR),
	\item continuum fit parameters, and 
	\item a postcard per observation.
\end{itemize}
\vspace{-6pt}
In addition, the SPIRE Automated Feature Extraction CATalogue (SAFECAT) provides a combined catalogue of both {\it sparse} and {\it mapping} results. All products are provided as multi-extension FITS tables. In this section we describe each of these products in detail.
\vspace{-12pt}
\subsection{Significant feature catalogues per observation}\label{sec:obsCat}
Significant feature catalogues per observation contain: the measured frequency of features with $|$SNR$|$ $\ge$ 5, the error on the measured frequency; the SNR measured using the fitted peak and the local noise in the residual spectrum after the total fitted model has been subtracted; which detector or array the feature was found in, (i.e.,  SLWC3 or SSWD4 for sparse-mode observations and SLW or SSW detector array and row, column spaxel for mapping observations); a feature flag (\S\,\ref{sec:featureFlags}); and a column indicating if the feature was detected or modified by the neutral carbon check (\S\,\ref{sec:neutralC}).

The catalogue per observation provides meta data with information on the observation, such as observation identifier, the target name, the operational day, source coordinates, spatial sampling mode, spectral resolution mode, bias mode, calibration type, and source extent. Additionally, the meta data provides information on FF input and results such as identifiers to show whether HPDPs were used (\S\,\ref{sec:HPDPs}); the masking width at the edge of the bands, definitions for feature flags, maximum continuum flux, and number of features found separately in the SLW and SSW band. For sparse observations, the meta data includes an estimate of the source radial velocity, associated error, and a flag indicating how the velocity estimate was derived \citepalias{FFredshift}. For mapping observations, the velocity estimate meta data are presented as tables in separate FITS extensions, with each element corresponding to a spaxel in an SLW/SSW combined projection grid (see \S\,\ref{sec:redshift} and \citetalias{FFredshift}). The number of features in each element of the combined grid is also presented in its own extension table.
\vspace{-12pt}
\subsection{SAFECAT}\label{sec:combined}
%
There is a combined significant feature catalogue denoted as SAFECAT that contains a row for every feature detected by the FF with $|$SNR$|$ $\ge$ 5, and includes all sparse and mapping observations. SAFECAT contains several columns providing information about the detected features and allowing for diverse search criteria. Information in these columns can be organized into four rough groups: observation, observing mode and calibration, input spectrum, and FF outputs. Observation information includes observation identifier, and operation date. Observing mode and calibration columns include source extent (i.e., point cf. extended), calibration mode, spatial sampling mode, and whether HPDPs were used. Input spectrum information includes which band/detector array the feature was detected in, the row and column of the corresponding spaxel (-1 for sparse observations), coordinates of the spaxel (mapping) or nominal pointing (sparse). The FF output columns include fitted feature frequency and associated error, SNR of the fitted peak to local noise in the full residual, feature flag, velocity estimate and associated error, velocity flag indicating which method was used, and a neutral carbon flag indicating features detected or modified by the neutral carbon check routine. 

In total 167,525 features are included in the SAFECAT. Tab.\,\ref{tab:detectionStats} presents a more detailed outline of where these features were detected. Some of these features are duplicate in that there is both point-source and extended calibration FF data for some of the sparse observations. 
With 1041 HR sparse observations, 868 are input to the FF line search (excluded targets are mentioned in \S\,\ref{sec:input}) and have continuum fit parameters reported in SAFECAT. Of the 643 HR sparse observations with spectral features identified by the FF, 287 are included only in the point-source calibrated data and 356 have entries in both the point and extended calibration FF data.
\begin{table}
%
\begingroup
\begin{center}
\newdimen\tblskip \tblskip=5pt
\caption{\label{tab:detectionStats}The number of features detected by the FF broken down into various categories. The first column splits the results into the sparse and mapping data sets. The second column indicates the data calibration mode (point or extended), followed by the third column which indicates the source classification, point-like, semi-extended, or extended. The `off-axis' row includes the search results from the off-axis detectors within HR sparse observations (see \citetalias{FFlineID}). The fourth column gives the number of observations with detected features. The fifth and sixth columns indicate the number of features detected in the SLW and SSW detector arrays, respectively.}
\nointerlineskip
\small
%
\newdimen\digitwidth
\setbox0=\hbox{\rm 0}
\digitwidth=\wd0
\catcode`*=\active
\def*{\kern\digitwidth}
\newdimen\signwidth
\setbox0=\hbox{+}
\signwidth=\wd 0
\catcode`!=\active
\def!{\kern\signwidth}
%
\tabskip=2em plus 2em minus 2em
\halign to \hsize{\hfil#&*#\hfil&\hfil#\hfil&\hfil#\hfil&\hfil#\hfil&\hfil#\hfil& \hfil#\hfil*&#\hfil\cr
 &\multispan6\hrulefill& \cr
\noalign{\vspace{-8.0pt}}
 &\multispan6\hrulefill& \cr
 & FF& Data& Source& Obs.& SLW& SSW& \cr
 & Group& Cal.& Class.& Count& Lines& Lines& \cr
\noalign{\vskip 2.5pt}
\noalign{\vspace{-8pt}}
 &\multispan6\hrulefill& \cr

& HR Sparse& Pt. source&  all&  643& *6\,299& **5\,606& \cr
&       &           &     pt.&  285& *3\,547& **3\,874& \cr
&       &           &   semi.&  243& *2\,010& **1\,355& \cr
&       &           &    ext.&  115& **\,742& ***\,377& \cr
\noalign{\vspace{-6pt}}
& & \multispan5\hrulefill& \cr
& & Ext. source& all&           348& *2\,564& **1\,538& \cr
&       &    & semi.&           232& *1\,844& **1\,175& \cr
&       &     & ext.&           116& **\,720& ***\,363& \cr
\noalign{\vspace{-6pt}}
& & \multispan5\hrulefill& \cr
& & & off-axis&           509& 17\,868& *12\,852& \cr
\noalign{\vspace{-6pt}}
& \multispan6\hrulefill& \cr
& HR Mapping& Ext. source& all& 205& 99\,784& 141\,368& \cr
\noalign{\vskip -5.5pt}
 &\multispan6\hrulefill& \cr
\noalign{\vspace{-10pt}}
\noalign{\vskip 2.5pt}
}
\end{center}
\endgroup
\vspace{-18pt}
\end{table}
\vspace{-12pt}
\subsection{Continuum fit parameters}\label{sec:continuum}
For each observation the best fit to the continuum, $P(\nu)$, is provided in the form of 3$^{rd}$ order polynomial parameters (p$_i$) of the form
\vspace{-6pt}
\begin{equation}
\label{eq:cont}
P(\nu) = {\rm p}_0 + ({\rm p}_1\,\times\,\nu) + ({\rm p}_2\,\times\,\nu^2) + ({\rm p}_3\,\times\,\nu^3),
\end{equation}
where $\nu$ is the frequency.  
A second order polynomial is used for LR observations.

Each fitted continuum parameters file contains a table of the parameters obtained from fitting the continuum of the spectrum, and the associated error, from the centre detectors (for sparse observations) or each spaxel of the SLW and SSW cubes (for mapping observations). For sparse observations the metadata reports the observation identifier and the nominal pointing coordinates. For mapping observations there are additional columns in the table for spaxel coordinates both in terms of spaxel row and column indices, and RA and DEC. 
%
%
\vspace{-12pt}
\subsection{The Feature Finder postcards}\label{sec:postcards}
\begin{figure}
\centering
\includegraphics[trim = 0mm 0mm 0mm 0mm, clip,width=\columnwidth]{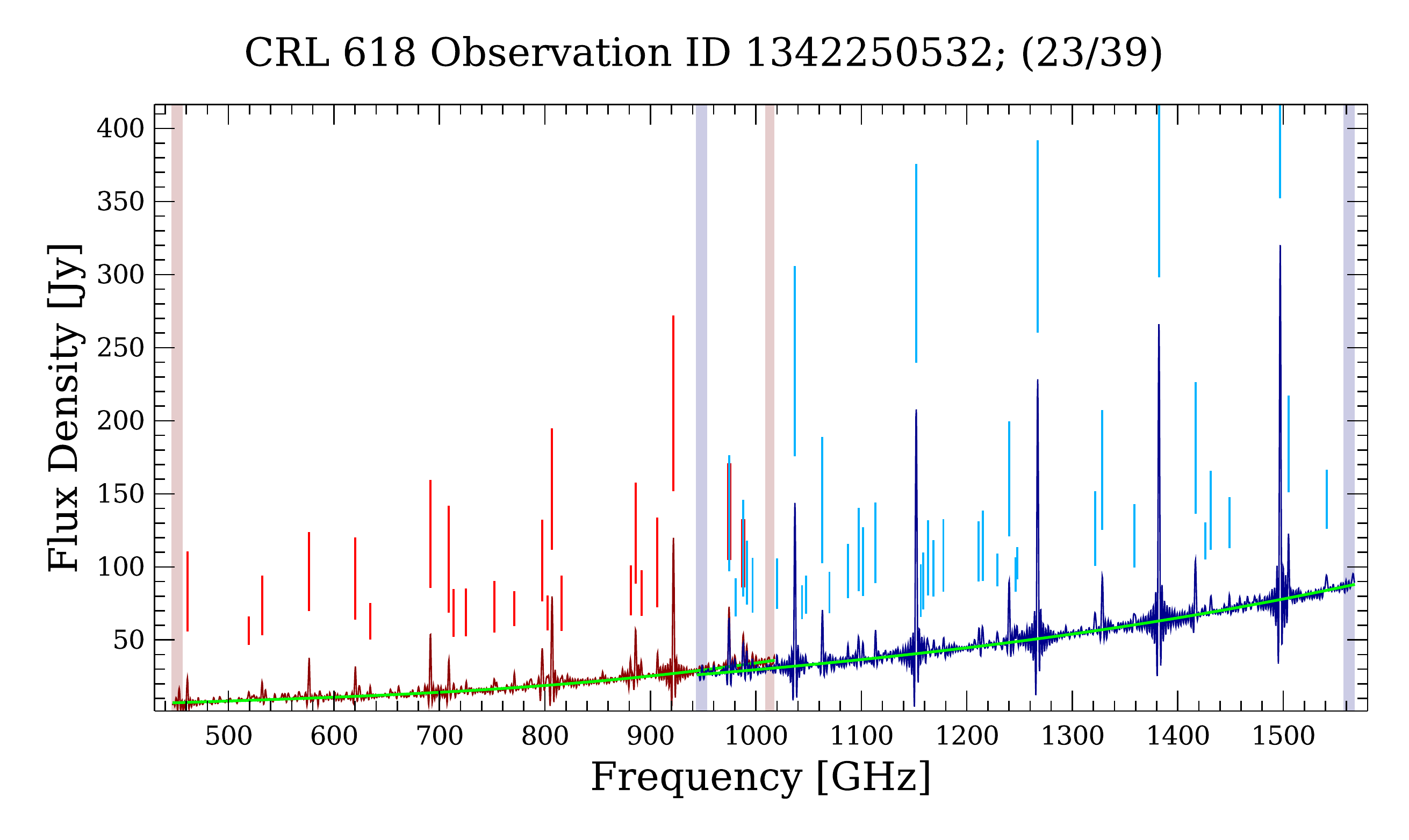}
\includegraphics[trim = 0mm 0mm 0mm 0mm, clip,width=\columnwidth]{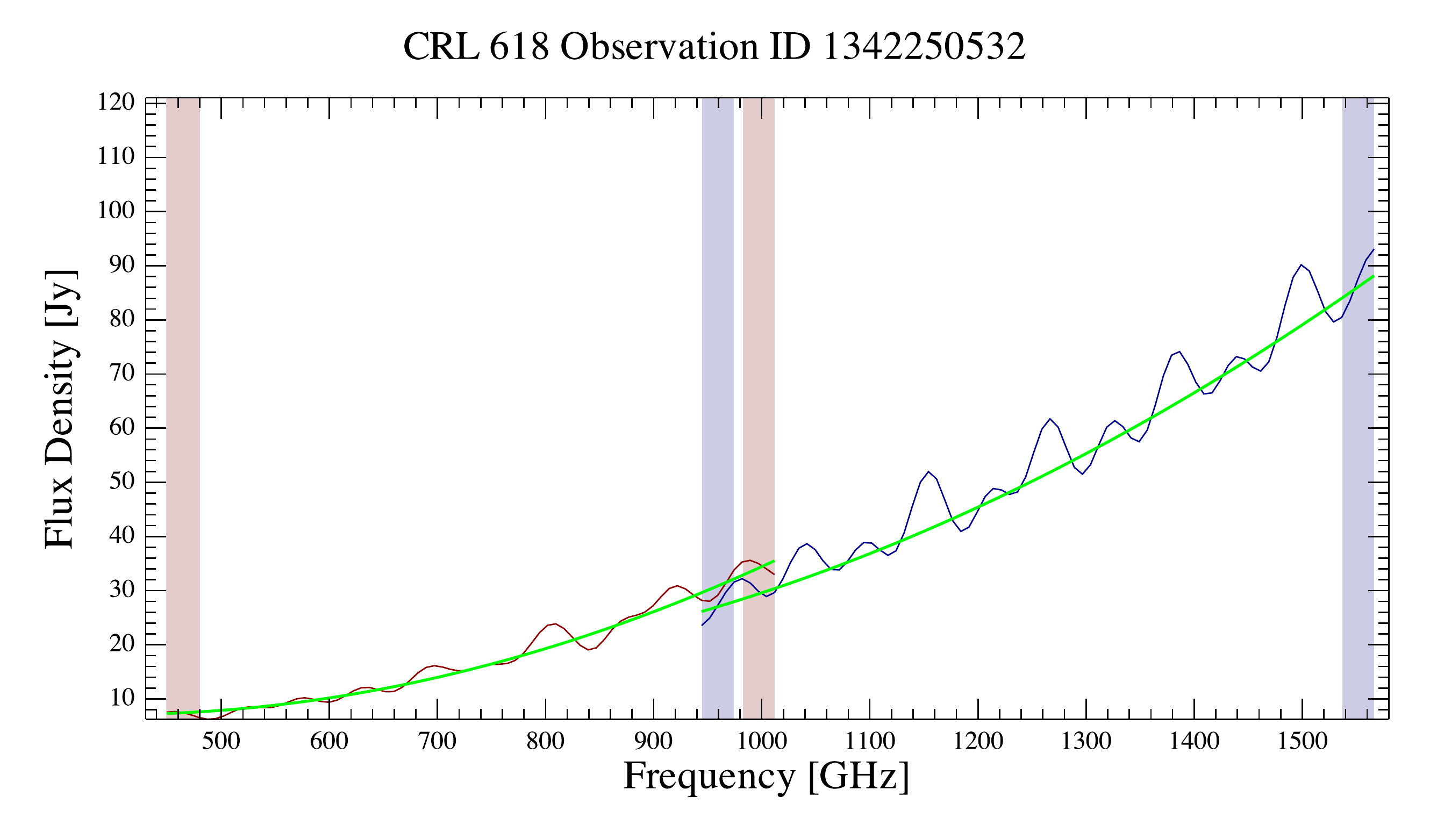}
\vspace{-18pt}
\caption{\label{fig:postcardSparse}An example of an HR sparse FF postcard (top) and a LR postcard (bottom) for an observation taken in H+LR mode. The HR spectrum shows a rich compliment of spectral features, heavily smoothed out in the LR version. The SLW and SSW band edge regions are outlined by full-height vertical bars.  The polynomial continuum fit is shown in green. The vertical sticks in the upper figure represent features identified by the FF, with height proportional to the SNR of the identified feature.}
\vspace{-12pt}
\end{figure}

The FF postcards are provided for quick inspection and qualitative analysis of FF results. An example postcard for sparse mode is shown in Fig.\,\ref{fig:postcardSparse}, upper panel: it presents the input spectra, the best fit to the continuum, and vertical bars representing the features found with their height scaled proportionate to the SNR. Only the spectra from the central detectors of the two bolometer arrays (SLWC3 and SSWD4) are presented, since these are the only spectra processed for these observations\footnote{Only two science and a few calibration sparse observations had the central detectors pointed directly off-source.}. For sparse-mode observations the data is calibrated in flux density units of Jy.

For LR observations, continuum parameters are included in the catalogue, however no features are provided. LR mode is intended to provide a continuum measurement and preliminary testing indicates that there are few features found in these observations. Two postcards are provided for LR, however, as many targets are semi-extended in nature and simply by comparing the point- and extended-calibration postcards it may be possible to gauge which calibration scheme is more appropriate or if further processing is required to correct for partial extent (see SDRG). Fig.\,\ref{fig:postcardSparse}, bottom panel shows an example of an FF postcard for the LR part of an H+LR observation. Only the continuum curves for LR are stored, and there are no features identified, although as in the example, there could be features identified in the counterpart HR spectrum. 

\begin{figure*}
\centering
\includegraphics[trim = 0mm 0mm 0mm 0mm, clip,width=1.0\textwidth]{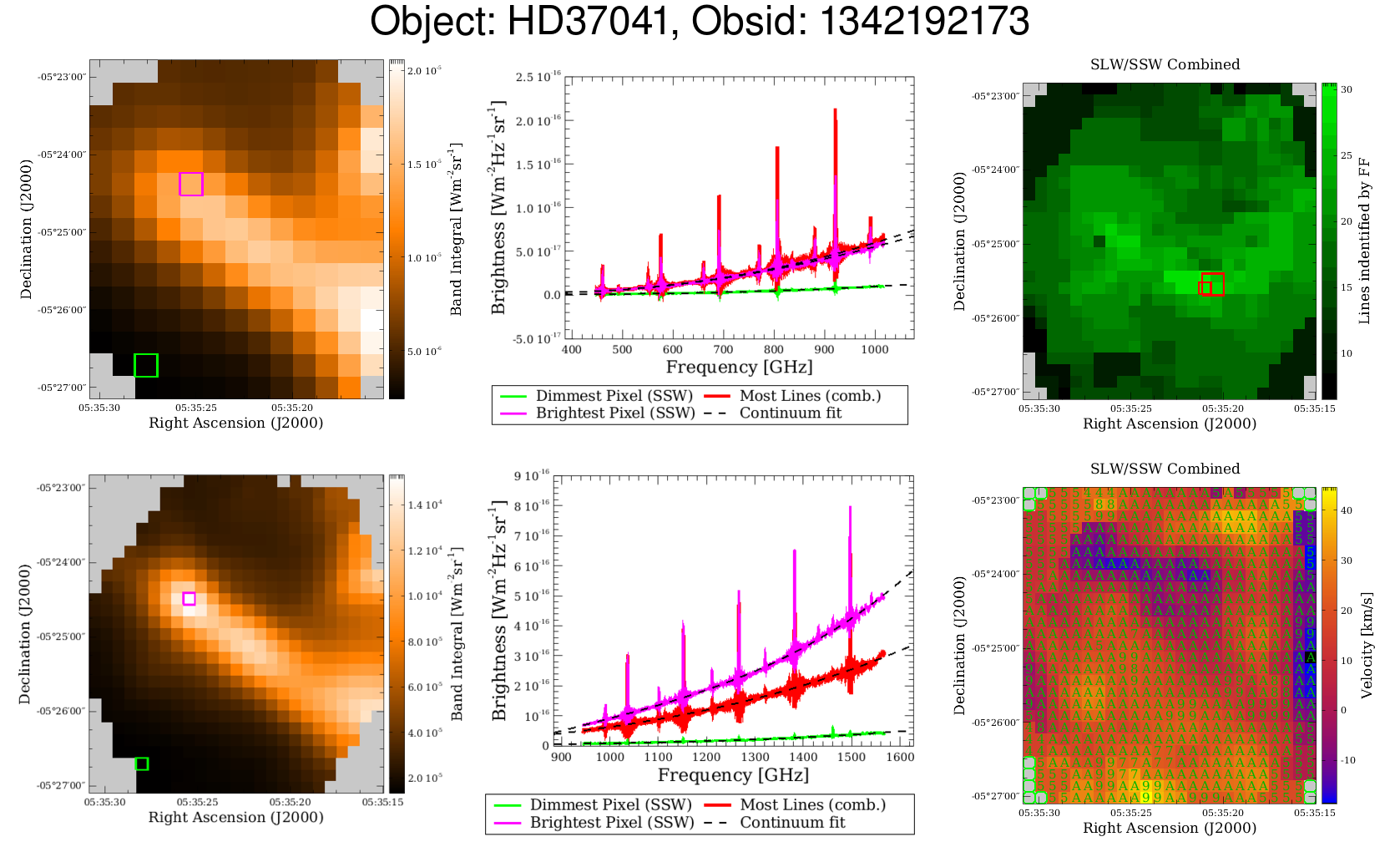}
\vspace{-18pt}
\caption{A sample mapping postcard. The first column shows the total integrated flux of the SLW (top) and SSW (bottom) cubes. The second column presents spectra corresponding to the highest SSW integrated flux, lowest SSW integrated flux, and most identified features (highlighted by squares with the same color in the adjacent plots), plotted in magenta, green, and red, respectively. The top SLW plot shows the closest associated SLW pixel for the corresponding brightest/dimmest SSW spectra. The third column presents results for the SLW/SSW combined projection grid with the top figure showing the number of lines projected into each grid element and the bottom figure showing the resulting velocity map.}
\label{fig:postcardMapping}
\vspace{-12pt}
\end{figure*}

Mapping postcards consist of a composite of 6 panels as shown in Fig.\,\ref{fig:postcardMapping}. The first column presents the total integrated flux maps of the SLW (top) and SSW (bottom) cubes. Magenta and green squares highlight the spaxels with the highest and lowest integrated flux, respectively, from the overlapping region of the two cubes. Spectra from these spaxels and their fitted continua are presented in the adjacent plots in the second column. The third column shows results for the SLW/SSW combined projection grid (see \S\,\ref{sec:redshift}). The top-right figure indicates the number of features projected onto each grid element with the red square highlighting the element with the most features. The corresponding SLW and SSW spectra and their fitted continua are plotted on the top and bottom figures in the centre column, respectively. The bottom-right figure shows the resultant velocity map (see \S\,\ref{sec:redshift}). Each grid element is annotated to indicate the spectral features used to derive the associated velocity estimate. Integers ranging from 3 to 9 indicated how many $^{12}$CO rotational feature candidates were identified and used to derive the estimate. The letter ``A'' is presented when all 10 nominal CO rotational emission features were used. When [NII] emission is used, the annotation is ``N''. For mapping observations, the data are calibrated in flux density units (i.e., W\,m$^{-2}$\,Hz$^{-1}$\,sr$^{-1}$).

\vspace{-12pt}
\section{Feature Finder Validation \& Results}\label{sec:FFvalid}
This section presents the methods used to validate the core FF routines and output products, including presentation of the validation methods, and the presentation of some of the key FF product results. Validation for the radial velocity estimating routines (\S\,\ref{sec:redshift}) and neutral carbon check(\S\,\ref{sec:neutralC}) is provided in the companion papers \citetalias{FFredshift} and \citetalias{FFncc}, respectively.
%
\vspace{-12pt}
\subsection{Completeness}\label{sec:completeness}
%
The completeness of the FF catalogues represents the fraction of real features detected, in the context of our testing framework, as a function of SNR. The completeness was estimated using Monte Carlo simulations. Repeated observations of the SPIRE dark sky field (RA:17h40m12s and Dec:+69d00m00s (J2000)) provide realistic noise levels, so these were used as a base population for simulating test spectra. There is an asymmetry present in the SPIRE FTS ILS that leads to a shortfall in measured line flux of 2.72\%, when calculated from fitted sinc parameters (see \citetalias{Hopwood15}). To better replicate the fitting performed by the FF it is important to introduce this asymmetry into the simulated spectra. Therefore spectral lines were added using an empirical line profile.

Several completeness test iterations were run. For each test a dark sky observation was selected at random, from a parent pool of all deep ($>$\,15 repetitions) high-resolution dark sky observations. Both the spectra from the SLW and SSW centre detectors were used. For the first two test sequences, random lines were added to the base spectrum, with ten lines added for the first test and four for the second. As the majority of SPIRE FTS observations are of targets that exhibit some level of $^{12}$CO emission, a template was created to include a $^{12}$CO ladder for the third test sequence. 

The template was based on co-added observations of NGC\,7027  (see \citetalias{Hopwood15} for details on how the data were co-added and cleaned of faint lines). To derive the template, a sinc profile per $^{12}$CO line and a 3$^{\rm rd}$ order polynomial were simultaneously fitted to the co-added data. The best fit sinc-parameters were used to add copies of the SPIRE FTS empirical line profile at the position of the $^{12}$CO lines. The $^{12}$CO lines of NGC\,7027 are unusually strong and suffer from less blending than spectral lines in most other SPIRE FTS observations. Therefore, to represent more typical spectra, the $^{12}$CO template was scaled to 30\%, with the continuum scaled to 60\%. To ensure the simulated spectra covered a range of line strengths and source velocities, two additional modifications are imposed. First, the $^{12}$CO ladder was scaled by a flux factor randomly generated from a uniform distribution between 0.1 and 1.0. This resulted in the simulated spectra having SSW flux levels that varied from a few Jy to around 90\,Jy. Second, to introduce a range of source velocities in the simulated population, a random redshift was generated from a Gaussian distribution with a mean of zero and $\sigma$ of 0.025. The $^{12}$CO ladder was redshifted and normalised using the associated luminosity distance. Finally, for each band of each simulated spectrum, four lines were added at random frequencies, with randomly generated SNRs. For all three tests, the random lines were allowed a maximum SNR of 100, and 20,000 simulated spectra were generated. All spectra were run with the full FF process and the results assessed to see how many of the features were found. A line was deemed successfully detected if there was a feature in the output that was within 2\,GHz of the simulated input frequency.\\

The completeness is calculated as the fraction of lines recovered per SNR bin, with associated errors of 
\begin{equation}
\label{eq:complete}
\sigma_C = \sqrt{\frac{C(1 - C)}{{\rm N}_{\rm Total}}},
\end{equation}
where {\it C} is the completeness , ${\rm N}_{\rm Total}$ is the total number of added lines, and $\sigma_C$ is the associated error. 

\begin{figure*}
\centering
\makebox[\textwidth][l]{
\hspace{-3pt}\includegraphics[trim = 0mm 0mm 0mm 0mm, clip,width=0.33\textwidth]{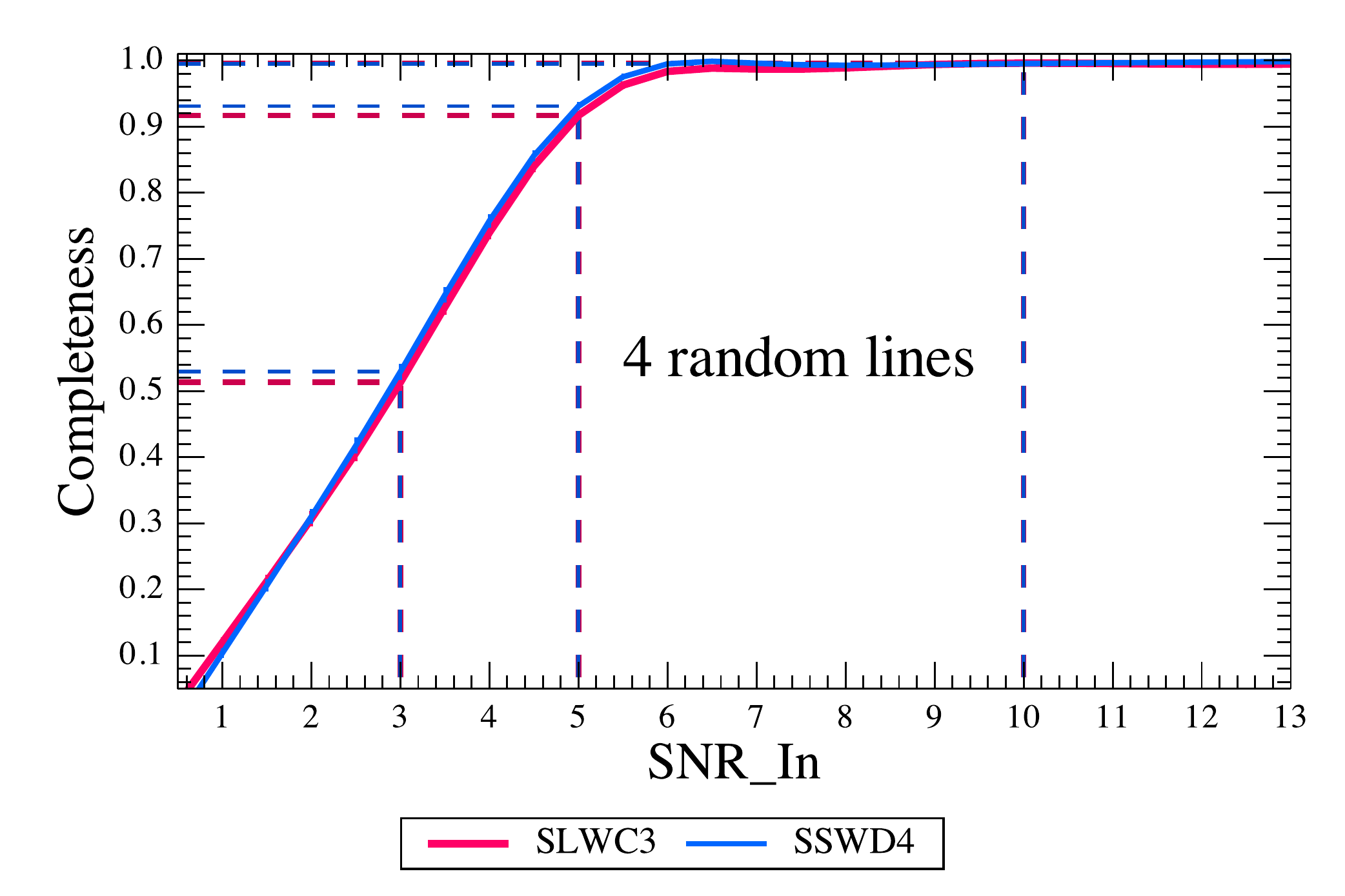} 
\includegraphics[trim = 0mm 0mm 0mm 0mm, clip,width=0.33\textwidth]{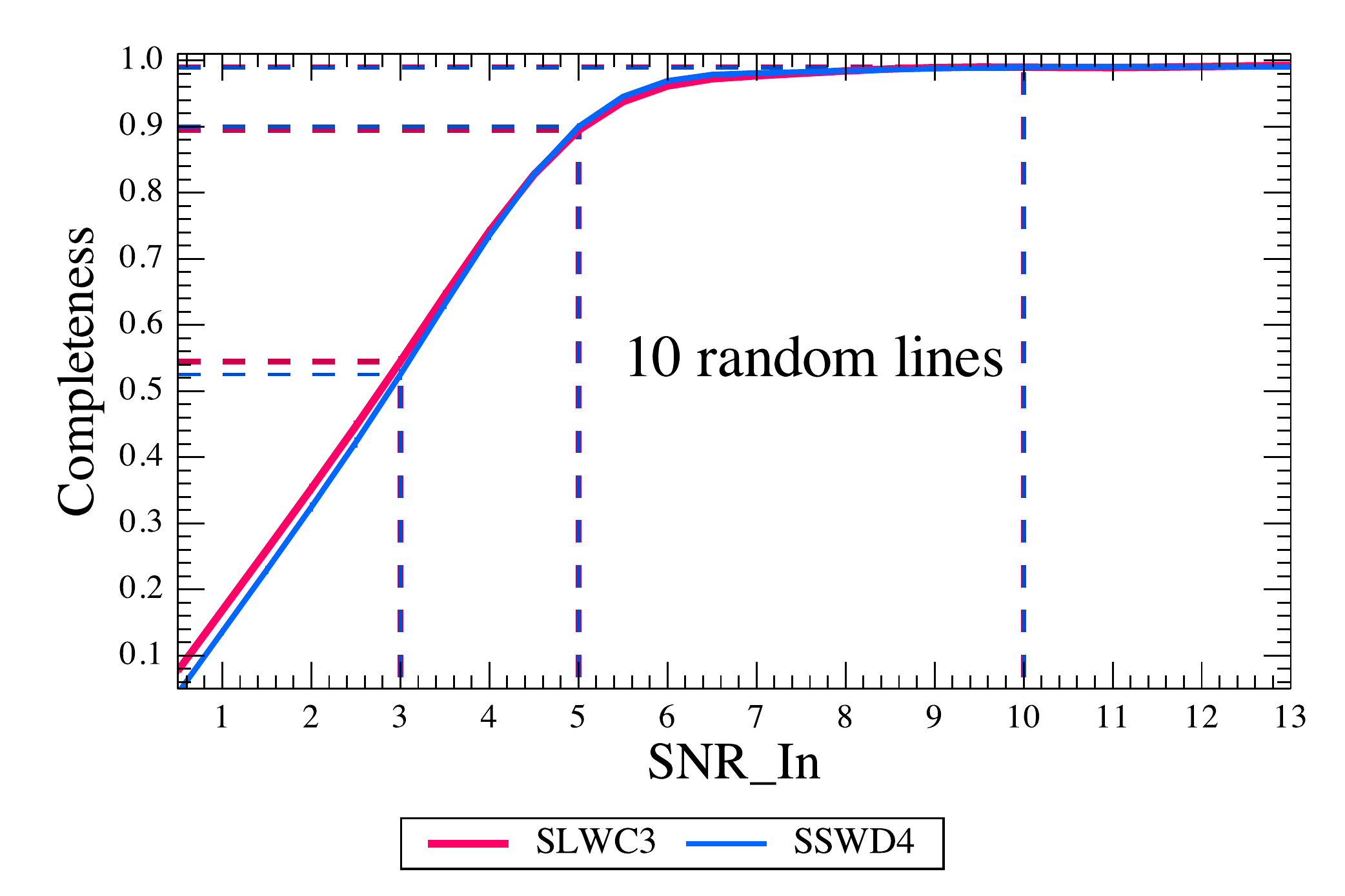} 
\includegraphics[trim = 0mm 0mm 0mm 0mm, clip,width=0.33\textwidth]{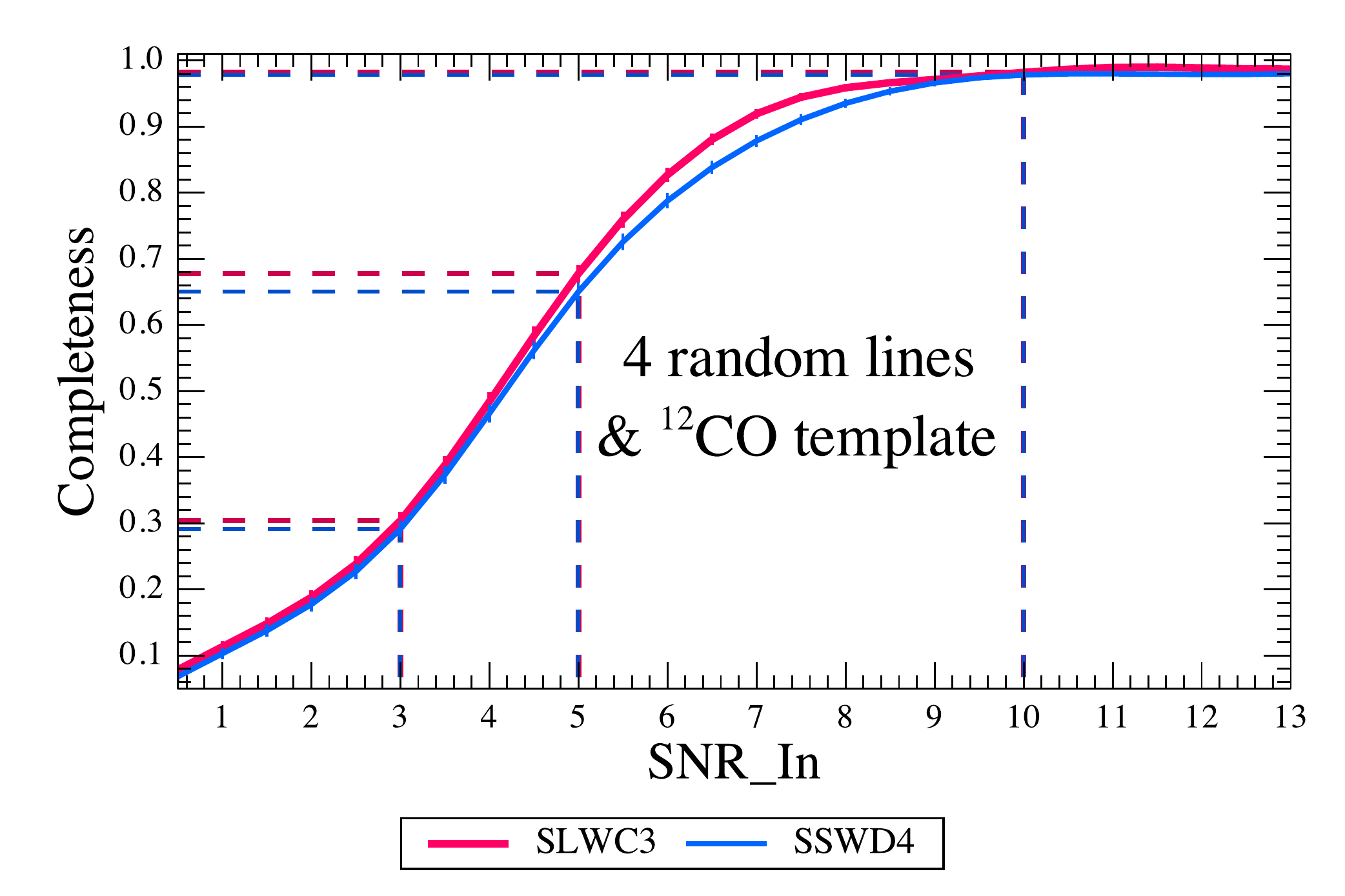}
}
\vspace{-18pt}
\caption{FF completeness as a function of input SNR. The left and middle panels show results for random line injection only, whereas the results in the right panel includes the addition of a $^{12}$CO ladder. For all three tests, the completeness is approximately 100\% for features with SNR greater than 10\,$\sigma$. However, for weaker features, the completeness depends on how rich in spectral features the spectrum is and the range of strengths of these features. For instance, at 5\,$\sigma$ the completeness drops from an average of 93\% when there are only four features added to each simulated spectra (the left case) to an average of 68\% if there is a significant $^{12}$CO ladder. The drop in completeness reflects the change in noise properties of SPIRE FTS spectra, which are dominated by the systematic noise of the sinc wings for sources that exhibit a spectrum rich in lines.}
\label{fig:completeness}
\vspace{-12pt}
\end{figure*}

Due to the large statistical sample number for each test, the associated errors are small. Fig.\,\ref{fig:completeness} shows the completeness results for the SLW and SSW detectors for all three tests. The completeness for 3, 5 and 10\,$\sigma$ are given in Tab.\,\ref{tab:completeness}. From Fig.\,\ref{fig:completeness} and Tab.\,\ref{tab:completeness}, the FF catalogues are essentially complete for spectral features with SNRs greater than 10. Below this threshold the completeness is affected by how rich in features a particular spectrum is and how strong those features are. If there are many strong features, such as for test 3 that includes a $^{12}$CO ladder, the completeness at 5\,$\sigma$ is approximately 20\% lower than for spectra containing only four random lines. The change in completeness reflects the different noise properties of SPIRE FTS data depending on the source configuration. When there are only a few weak spectral lines present, the noise is predominantly random. When there are many spectral lines present, the noise may be dominated by the systematic noise introduced by the wings of the lines themselves. If the blending of sinc wings is significant, then faint features are more easily lost and spurious faint features are more easily detected.  Validation of the FF routine with respect to false positives and the low SNR cutoffs was conducted using the lineless calibration sources as presented in \S\,\ref{sec:featureFlags} and Fig.\,\ref{fig:noisy}.
\begin{table}
\caption{Completeness results for the three Monte Carlo simulations. For test 1, four random lines were added to each simulated spectrum. For test 2, ten random lines were added. For test 3, four random lines and a $^{12}$CO ladder were added. The 3\,$\sigma$, 5\,$\sigma$ and 10\,$\sigma$ completeness (C) is given for each test as C$_{\rm SLWC3}$/C$_{\rm SSWD4}$.}
\vspace{-12pt}
\begin{center}
\begin{tabular}{lccc}
\hline\hline
Test & 3\,$\sigma$ & 5\,$\sigma$ & 10\,$\sigma$ \\ \hline
1 &  0.51/0.53 & 0.92/0.93 & 1.00/1.00 \\
2 &  0.55/0.53 & 0.90/0.91 & 0.99/0.99 \\ 
3 &  0.30/0.29 & 0.68/0.65 & 0.98/0.98 \\ \hline 
\end{tabular}
\end{center}
\label{tab:completeness}
\vspace{-12pt}
\end{table}
%
%
\vspace{-12pt}
\subsection{Comparison with ``Python Peaks''}\label{sec:pythonPeaks}
The FF method relies solely on the standard data processing of SPIRE FTS data.  Occasionally, there are unique conditions in the observation data causing failure of the FF algorithm.

To help overcome this challenge, and to validate the standard FF results, a search for spectral features has been developed outside of HIPE, using Python \textsc{SciPy} methods in the \textsc{scipy.signal} package \citep{scipy}. The continuum is subtracted from the SPIRE FTS apodized data \citep[][]{Fulton14a}, using the best fit parameters already determined by the FF. A wavelet function, as implemented in \textsc{scipy.signal.find\_peaks\_cwt}, is used to search for peaks with SNR\,$>$\,8. The noise can then be estimated with these strong peaks omitted. A second search for peaks is then carried out with the \textsc{SciPy} method \textsc{scipy.signal.argrelextrema}, using the noise estimate in order to find local maxima with SNR\,$>$\,3 (increased to SNR\,$>$\,5 in the overlap region of the two FTS bands). We call this two-step alternative method to find peaks in apodized SPIRE FTS data ``Python Peaks'' (PP). It should be noted that an SNR of 3 calculated by PP approximately corresponds to an SNR of 5 for the final FF SNR estimates due to the different noise properties in the apodized and raw un-apodized spectra. The PP results are compared to the nominal FF features and a spectroscopic infrared line template list. This comparison is used to improve the FF results by adding missed features and removing spurious ones. 
%

As a consequence of using apodized data, PP suffers less from wing-fitting and false detections in areas of high noise, such as the fringing for spectra of partially-extended sources. Additionally, due to the necessary checks during the FF, there are some cases of significant features being wrongly discarded. These lost features can be restored using the PP/FF/line-list comparison. As all SPIRE FTS spectra of partially-extended sources have been flagged as such, the PP/FF comparison can also be used to identify if there is a fringe-fitting issue, as in these cases there is a large discrepancy in features to peaks found. If there are many more features in the initial FF results beyond 900\,GHz (in SLW), this indicates a problem and all features at frequencies greater than the rest frame frequency of $^{12}$CO(8-7) (i.e., 921.80\,GHz) are removed from the catalogue. Further details of this method, and the corresponding spectral line identification are available in \citetalias{FFlineID}.

%
%
\vspace{-15pt}
\subsection{Comparison with the SPIRE FTS calibration sources}\label{sec:calibrators}
\begin{figure}
\centering
\includegraphics[trim = 0mm 0mm 0mm 3mm, clip,width=\columnwidth]{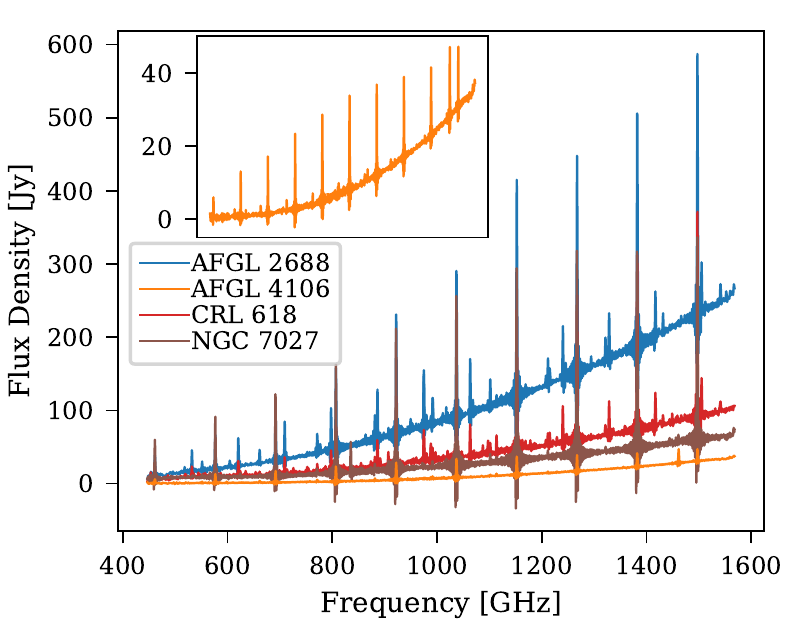}
\vspace{-18pt}
\caption{Example spectra for the four main SPIRE FTS line source calibrators \citepalias{Hopwood15}. The inset plot shows the AFGL\,4106 spectrum at a more suitable vertical plot range.}
\label{fig:egSpecLineSources}
\vspace{-21pt}
\end{figure}

The four sources AFGL\,2688, AFGL\,4106, CRL\,618, and NGC\,7027 are spectral-line rich, and were monitored throughout \textit{Hershel}'s mission as part of the SPIRE Instrument Control Centre's systematic programme of calibration observations \citep[][]{Swinyard2014}.\footnote{Spectral Calibrators DOI: \href{https://doi.org/10.5270/esa-xedr4yo}{doi.org/10.5270/esa-xedr4yo}}
A thorough analysis of the SPIRE FTS calibration source data is presented in \citetalias{Hopwood15}, which includes detailed line fitting. Example spectra for each of these sources are provided in Fig.\,\ref{fig:egSpecLineSources}, and show that these data cover a range in brightness, with SSW continuum levels between tens of Jy, for AFGL\,4106, to hundreds of Jy, for AFGL\,2688. There is also a difference between the sources in line strength and line density, which results in different levels of blending. In this section the line catalogues and SNRs from \citetalias{Hopwood15} are used as a comparison to the FF results for the observations of these four sources. Tab.\,\ref{tab:lineSources} provides information about the observation of these calibration sources.
All the observations included in the comparison were taken in HR sparse mode. The spectra used are corrected for pointing offset, 
and observations of AFGL\,4106 have a high level of background subtraction. 
The corrected data are available as HPDPs (\S\,\ref{sec:HPDPs}) via the  Spectrometer calibrators legacy data page.

\begin{table}
\vspace{-6pt}
\caption{Information on the SPIRE FTS calibration sources used for the FF comparison. The column ``N\_obs'' gives the number of observations recorded for the respective source.}
\vspace{-12pt}
\begin{center}
\hspace{-12pt}
\resizebox{0.4875\textwidth}{!}{%
\begin{tabular}{llllc}
\hline\hline
Name & Type & RA & Dec & N\_obs \\ \hline
AFGL\,2688& proto planetary nebulae& 21:02:18.78& $+$36:41:41.2& 23 \\ 
AFGL\,4106& post red supergiant& 10:23:19.47& $-$59:32:04.9& 30 \\ 
CRL\,618& proto planetary nebulae& 04:45:53.64& $+$36:06:53.4& 23 \\ 
NGC\,7027& planetary nebulae& 21:07:01.59& $+$42:14:10.2& 31 \\ 
\hline
\end{tabular}%
}
\end{center}
\label{tab:lineSources}
\vspace{-18pt}
\end{table}
%

%
\begin{figure*}
\centering
\makebox[\textwidth][l]{
\def\big{\hspace{-9pt}\includegraphics[width=\textwidth]{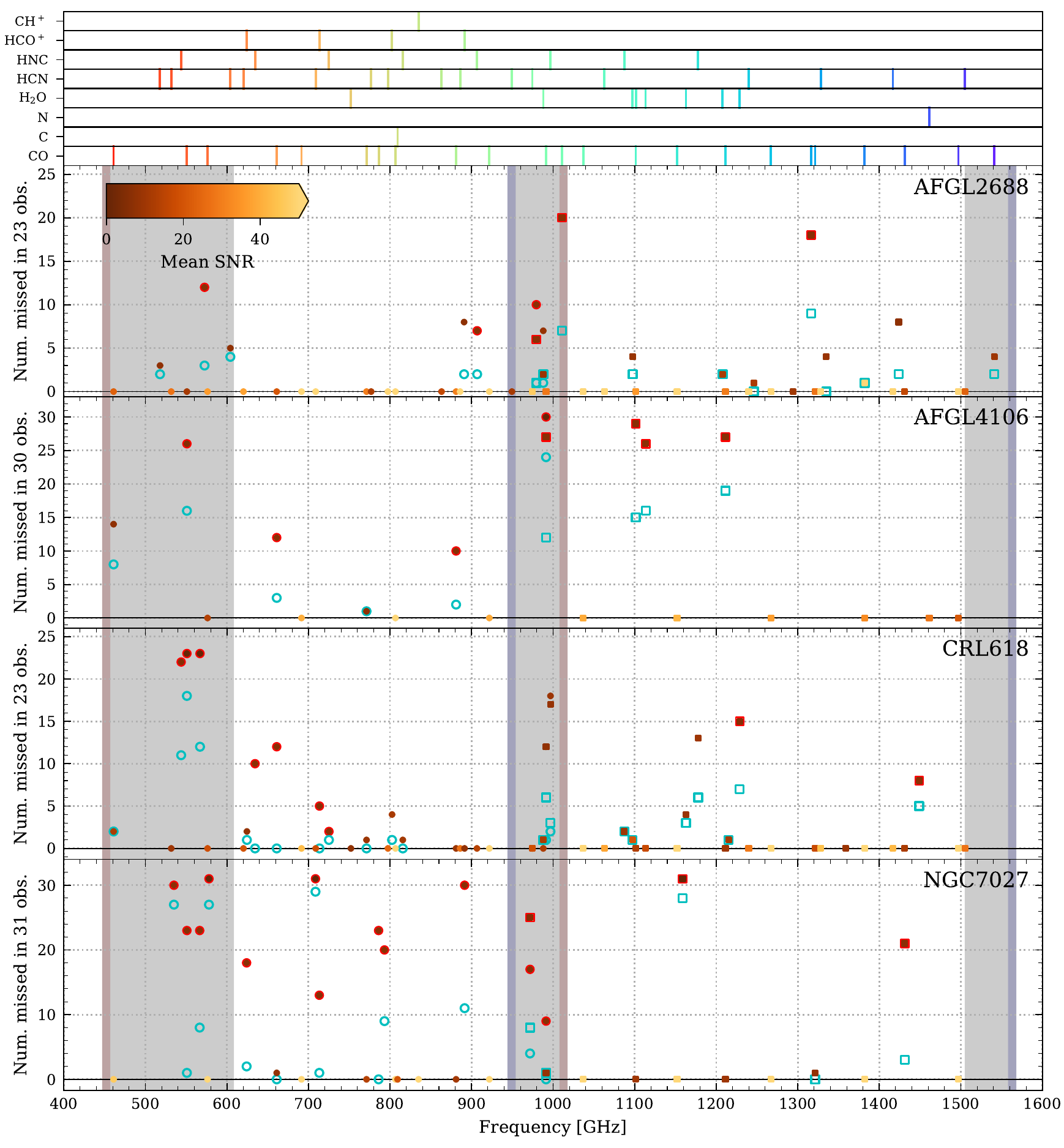}}
\def\lit{\includegraphics[trim = 0.0mm 0.0mm 0.0mm 0.0mm, clip, height=3.1cm]{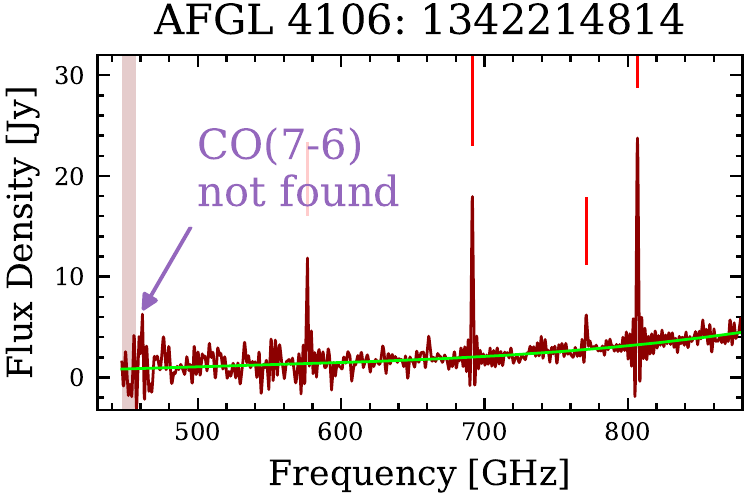}}
\stackinset{l}{338pt}{t}{203.5pt}{\lit}{\big}
%
}
\vspace{-12pt}
\caption{\label{fig:calibratorsHitMiss}Missed lines for the four SPIRE FTS calibration sources (see Fig.\,\ref{fig:egSpecLineSources}). Lines with SNR $>$ 5 for each observation were matched to \citetalias{Hopwood15},  
coloured by the mean SNR for a given transition. The SLWC3/SSWD4 features are shown with filled circles/squares, respectively, with a red outline added for features with mean SNR $<$ 7. A second match was performed using all found features, before the final SNR estimate and cut. The misses considering all features found are plotted as cyan circles/squares for SLWC3/SSWD4, and are only plotted where these are non-zero or lower than the number of missed lines in the final catalogues. Rest-frame line transition frequencies for identified species are provided at the top of the figure for reference (see \citetalias{FFlineID}). The noisy spectral regions are shaded, and the SLW and SSW band edges are marked with shaded vertical bars (as in Fig.\,\ref{fig:noisy}). 
The inset provides an example of a missing $^{12}$CO(4-3) line in 
AFGL\,4106 near the band edge within the noisy spectral region. 
}
\end{figure*}
%
Three cases were considered when assessing the FF results against the line catalogues: a line has been successfully found; a line has been missed; an extra feature has been found. If the assumption that all significant lines have been included in the line lists is employed, then in the latter case, the extra feature is likely to be spurious. However, where there are repeated extra features found at a similar frequency, it may be that a real line was not included in the line catalogues or that the line was measured with a SNR of less than 5. To compare the FF results with the known lines for each source, the following steps were taken per observation, while treating the results for SLWC3 and SSWD4 separately. 
\begin{itemize}
    \item The line lists were adjusted for each observation, so only lines with a measured SNR $\ge$ 5 were considered.
	\item Each remaining line position was compared to the features in the final FF catalogues and to all features found before final SNR estimate was taken and the SNR cut was applied.
	\item A feature detected using the FF within 0.5\,GHz of a feature in the reference catalogue was considered a match. If multiple features matched this criteria, the closest feature to the line position was taken. (Note: a 2\,GHz match window was used for the simulated feature validation in \S\,\ref{sec:completeness}.)
	\item If no feature was found to match, this was recorded as a missed line.
	\item After all line positions were checked, features in the FF catalogues without a match were recorded as extras.
\end{itemize}
%

\begin{figure*}
\centering
\includegraphics[trim = 0mm 0.0mm 0mm 0.0mm, clip,width=\textwidth]{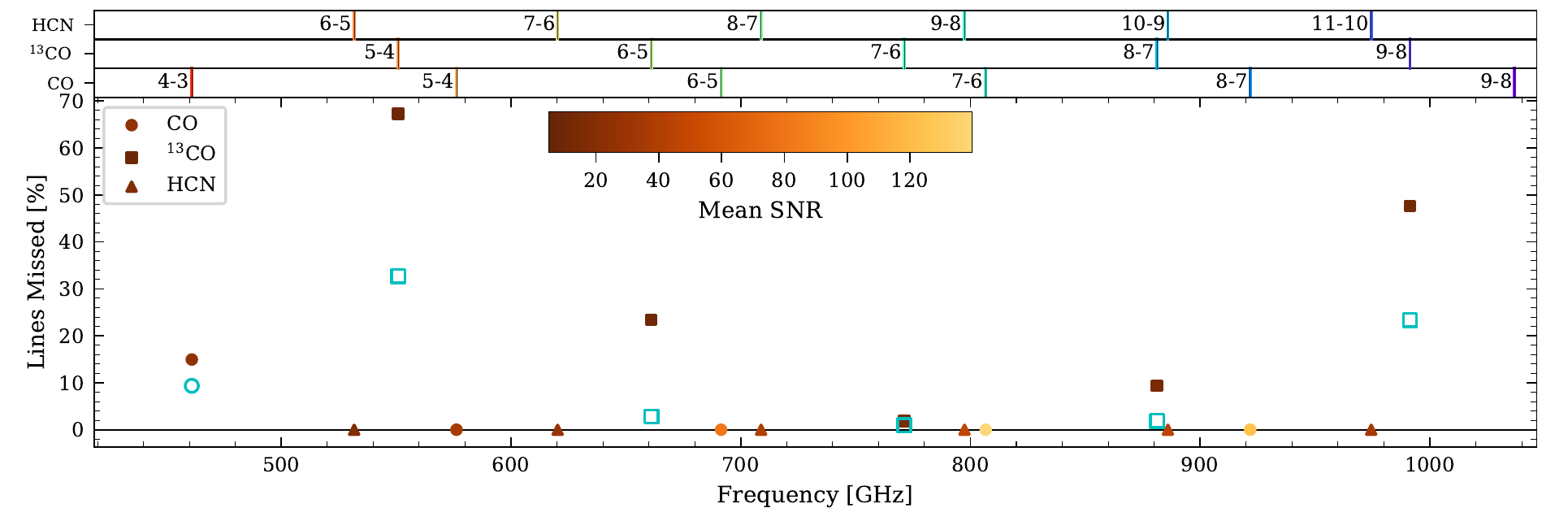}
\includegraphics[trim = 0mm 0.0mm 0mm 0.0mm, clip,width=\textwidth]{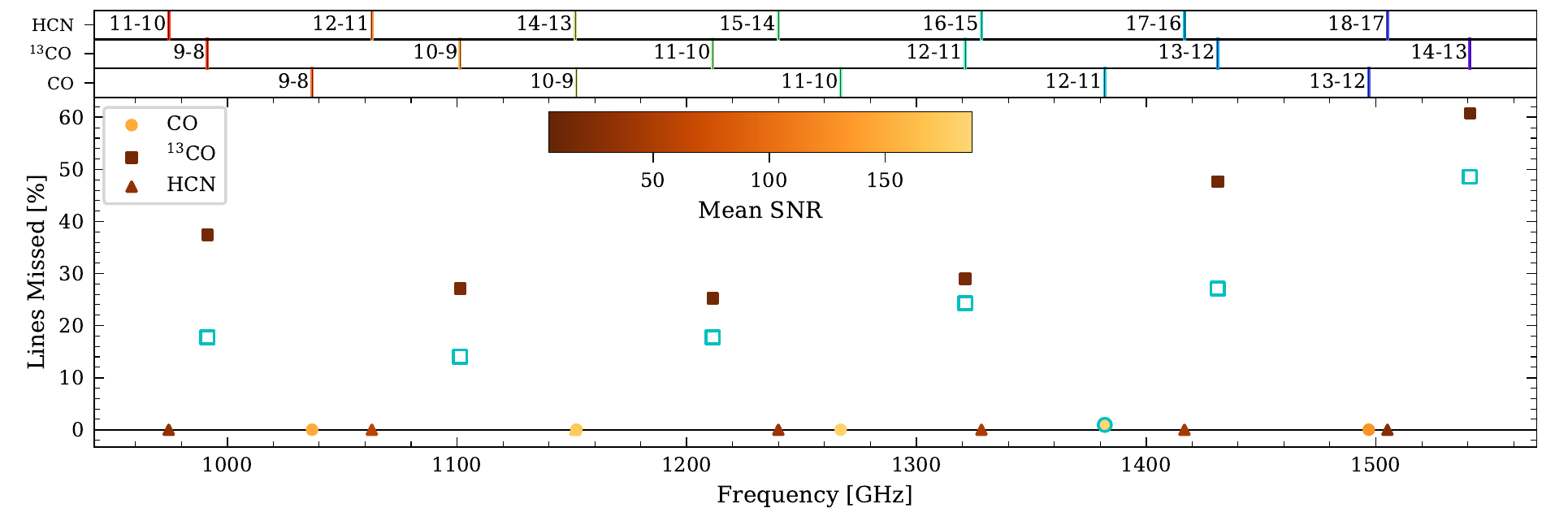}
\vspace{-15pt}
\caption{The number of spectral lines missed for rotational transitions of $^{12}$CO, $^{13}$CO and HCN within the SPIRE FTS frequency bands for the four main SPIRE FTS spectral line calibration sources. Each point gives the number of times a particular line was missed as a percentage of the total number, which are 96 for $^{12}$CO and $^{13}$CO, and 37 for HCN. Each point is coloured by the mean SNR for all lines included per detector. A second match was performed using all the features found, before the final SNR estimate and cut are applied. The number of misses when considering all features found are plotted as empty cyan symbols. The misses for all features are only plotted where these are non-zero or lower than the number of missed lines in the final catalogues. Features found in SLWC3 are shown in the top panel with those found in SSWD4 in the bottom panel. The line names and rest-frame transition frequencies are indicated at the top of each panel.}
\label{fig:speciesHitMiss}
\vspace{-12pt}
\end{figure*}

%
%
\vspace{-12pt}
For the four line source calibrators,  Tab.\,\ref{tab:statsPerSource1} collates the total number of lines matched (a hit), as well as the missed and extra features for each of the line sources. The overall success rate of the FF was assessed by considering the number of lines without a FF match (a miss) for all observations per source; false positives are assumed to be negligible for SNR\,$\ge$\,10 as discussed in \S\,\ref{sec:completeness}.
Fig.\,\ref{fig:calibratorsHitMiss}, presenting the same results as in Tab.\,\ref{tab:statsPerSource1}, shows the total number of misses for each line, compared to lines identified in \citetalias{Hopwood15}, while the symbols are coloured by the average SNR for that line, with a red outline if the average SNR $<$ 7. Also plotted (as cyan symbols) are the total number of misses when considering all features found (i.e., before the final SNR threshold cut is applied). Misses are observed mainly for lines near the band edges and within the noisy regions near the edges (see Fig.\,\ref{fig:noisy}), and features subject to high levels of blending. For all but NGC\,7027, a few lines at the ends of the SLW frequency band are missed for a higher proportion of the observations, although these all have low SNR, generally close to the lower limit of 5. The $^{12}$CO spectral features are rarely missed for these sources, with exceptions being 1 of 23 $^{12}$CO(12-11) lines (rest frequency 1\,382.00\,GHz) in AFGL\,2688, 12 of 23 $^{12}$CO(5-4) lines (rest frequency 576.27\,GHz) in AFGL\,2688, 14 of 30 $^{12}$CO(4-3) lines (rest frequency 461.04\,GHz) in AFGL\,4106, and 2 of 23 $^{12}$CO(4-3) lines in CRL\,618. All but one of these missed $^{12}$CO lines are within the noisy spectral regions. The majority of the missed features in the good spectral regions are associated with low SNR lines from less abundant isotopes/isotopologues (further details are provided in Fig.\,\ref{fig:speciesHitMiss}). 

\begin{table}
\begingroup
\begin{center}
\newdimen\tblskip \tblskip=5pt
\caption{\label{tab:statsPerSource1}For each of the four spectral calibration sources, the total numbers of FF features matched to a line (from \citetalias{Hopwood15}) is given in the `Hit' column. The `Miss: with SNR cutoff' and `Miss: no SNR cutoff' columns, present the total number of lines without a match when considering only those features above the SNR threshold applied by the FF (with SNR cutoff) or all features found prior to application of the SNR threshold (no SNR cutoff). The `Extra' column reports the total number of extra features found (i.e., no corresponding line in \citetalias{Hopwood15}). Results are given separately for the SLWC3 and SSWD4 detectors. For the numerical entries, the count is followed by a percentage, separated by a `/'. Each entry contains two sets of results, the upper entry reports data including the full detector band while the lower entry labelled `clean' presents the results within the lower-noise region of the band only (unshaded regions in Fig.\,\ref{fig:calibratorsHitMiss}).\vspace{-6pt}
}
\nointerlineskip
\small
%
\newdimen\digitwidth
\setbox0=\hbox{\rm 0}
\digitwidth=\wd0
\catcode`*=\active
\def*{\kern\digitwidth}
\newdimen\signwidth
\setbox0=\hbox{+}
\signwidth=\wd 0
\catcode`!=\active
\def!{\kern\signwidth}
%
\tabskip=2em plus 2em minus 2em
\halign to \hsize{*#\hfil&\hfil#\hfil& \hfil#\hfil& \hfil#\hfil& \hfil#\hfil& \hfil#\hfil*\cr
\noalign{\doubleline}
\noalign{\vskip -1.5pt}
      &     &    &      Miss:&      Miss:& \cr
Source& Band& Hit& with SNR&   no SNR& Extra\cr
      &     &    &   cutoff&   cutoff& \cr
\noalign{\vskip 4pt}
\noalign{\hrule}
\noalign{\vskip 4pt}
&      SLWC3& 583& *52*/**8.7\%& *15*/**2.5\% & *84\cr 
AFGL & clean& 431& *32*/**7.3\%& **6*/**1.3\% & *83\cr 
\noalign{\vskip -5pt}
 &\multispan5\hrulefill\cr
\noalign{\vskip -1pt}
2688 & SSWD4& 593& *70*/*11.3\%& *28*/**4.5\%& 168\cr 
& clean &     444& *38*/**8.2\%& *16*/**3.5\%& 159\cr 
\noalign{\vskip 2pt}
\noalign{\hrule}
\noalign{\vskip 3pt}
&      SLWC3& 246& *93*/*31.0\%& *54*/*18.0\%& **1\cr 
AFGL & clean& 180& *53*/*25.2\%& *30*/*14.3\%& **0\cr 
\noalign{\vskip -5pt}
 &\multispan5\hrulefill\cr
\noalign{\vskip -1pt}
4106 & SSWD4& 238& 109*/*36.3\%& *62*/*20.7\%& **2\cr 
 & clean&     220& *82*/*30.4\%& *50*/*18.5\%& **1\cr 
\noalign{\vskip 2pt}
\noalign{\hrule}
\noalign{\vskip 3pt}
& SLWC3&      618& 137*/*20.5\%& *49*/**7.3\%& *79\cr 
CRL & clean&  502& *49*/**9.7\%& **4*/**0.8\%& *77\cr 
\noalign{\vskip -5pt}
 &\multispan5\hrulefill\cr
\noalign{\vskip -1pt}
618 & SSWD4&  586& *74*/*11.9\%& *35*/**5.6\%& 160\cr 
& clean&      481& *44*/**8.7\%& *25*/**4.9\%& 120\cr 
\noalign{\vskip 2pt}
\noalign{\hrule}
\noalign{\vskip 3pt}
& SLWC3&      563& 269*/*39.4\%& 119*/*17.4\%& *43\cr 
NGC & clean&  440& 162*/*32.7\%& *56*/*11.3\%& *37\cr 
\noalign{\vskip -5pt}
 &\multispan5\hrulefill\cr
\noalign{\vskip -1pt}
7027 & SSWD4& 332& *79*/*21.2\%& *40*/*10.8\%& *83\cr 
& clean&      279& *53*/*17.1\%& *31*/*10.0\%& *82\cr 
\noalign{\vskip 1pt}
\noalign{\hrule}
\noalign{\vskip -5.5pt}
%
}
\end{center}
\endgroup
\vspace{-12pt}
\end{table}

In addition, AFGL\,4106 is embedded in high Galactic Cirrus, which leads to extra systematic noise, on both narrow and wide frequency scales, particularly towards the band edges. The inset within Fig.\,\ref{fig:calibratorsHitMiss} shows an example where  $^{12}$CO(4-3) is missed (in the AFGL\,4106 observation 1342214814), because the line intensity is at an amplitude similar to the local noise. One of the AFGL\,4106 observations (ID 1342208380) is excluded from our analysis due to an exceptionally large pointing error of $\sim$\,24 arcseconds. This results in the source being missed by SSWD4 and only partial source coverage with SLWC3.

\begin{figure}
\centering
\includegraphics[trim = 0mm 0mm 0mm 0mm, clip,width=\columnwidth]{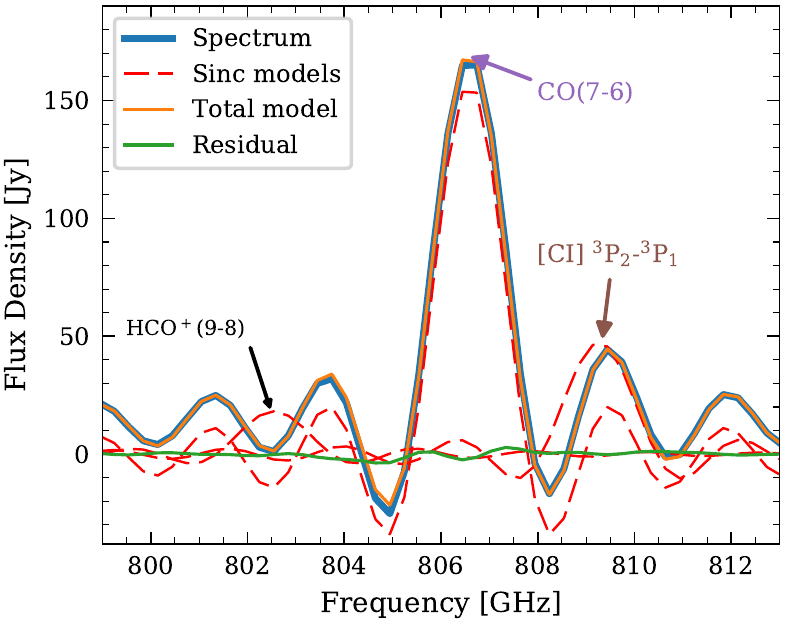}
\vspace{-18pt}
\caption{Sample NGC\,7027 spectrum demonstrating the close proximity of the $^{12}$CO(7-6) and [CI] features.}
\label{fig:CI_NGC7027}
\vspace{-12pt}
\end{figure}

Two lines of low significance, but not at the edge of the bands, are missed for NGC\,7027. However, these are in the wings of strong lines and likely rejected by the FF as potential ``wing fitting''. Thus, these are missed less often before the final FF checks. [CI]\,$\mathrm{^3P_2 - ^3P_1}$ is missed only once for all of the 31 NGC\,7027 observations, and is recovered using the neutral carbon check (see \S\,\ref{sec:neutralC}). Fig.\,\ref{fig:CI_NGC7027} illustrates the close proximity of the $^{12}$CO(7-6) and [CI] feature for NGC\,7027. 

Fig.\,\ref{fig:speciesHitMiss}, and Tabs.\,\ref{tab:speciesStats} and \ref{tab:rotStats} present the matched lines grouped by species and rotational transition. As explained above, only one $^{12}$CO line is missed in some AFGL\,4106 observations, and this does not change if all the features found are considered (plotted as cyan symbols, where different to the result when matching to the final FF catalogues). 
All but the lowest significance $^{13}$CO lines are matched for every observation, with the rate of missed features below 10\%. There are 12 HCN transitions within the SPIRE FTS bands that are detected with SNR $>$ 5, with no recorded misses. HCO+(7-6) has no misses when considering all the features found. It is, however, discarded as too faint or unreliable for 16\% of line source FF catalogue candidate features. HCO+(9-8) has several close (and more significant) neighbouring lines, including $^{12}$CO(7-6) and [CI]\,$\mathrm{^3P_2 - ^3P_1}$, so it is more often than not lost in the wings of these prominent features. However, the other lines in Tabs.\,\ref{tab:speciesStats} and \ref{tab:rotStats} are matched for most observations, regardless of their SNR, with few discarded by the FF SNR cut or reliability checks.\\
\begin{table}
\vspace{-3pt}
\caption{The mean SNR and percentage of times [CI]\,$\mathrm{^3P_2 - ^3P_1}$ and CH+(1-0) (rest frequency 835.08\,GHz) were missed for 31 observations of the line calibration sources.}
\vspace{-9pt}
\begin{center}
\begin{tabular}{lcccc}
\hline\hline
Species & N\_obs & $\nu$[GHz] & $\overline{\rm SNR}$ & \%miss \\ \hline
[CI]\,$\mathrm{^3P_2 - ^3P_1}$ & 31 & 809.13 & 24.8 & 3 \\
CH+(1-0) & 31 & 835.06 & 50.0 & 0 \\
\hline
\end{tabular}
\end{center}
\label{tab:speciesStats}
\vspace{-15pt}
\end{table}
\begin{figure*}
\centering
\makebox[\textwidth][l]{
\includegraphics[trim = 0mm 0mm 0mm 0mm, clip,width=0.33\textwidth]{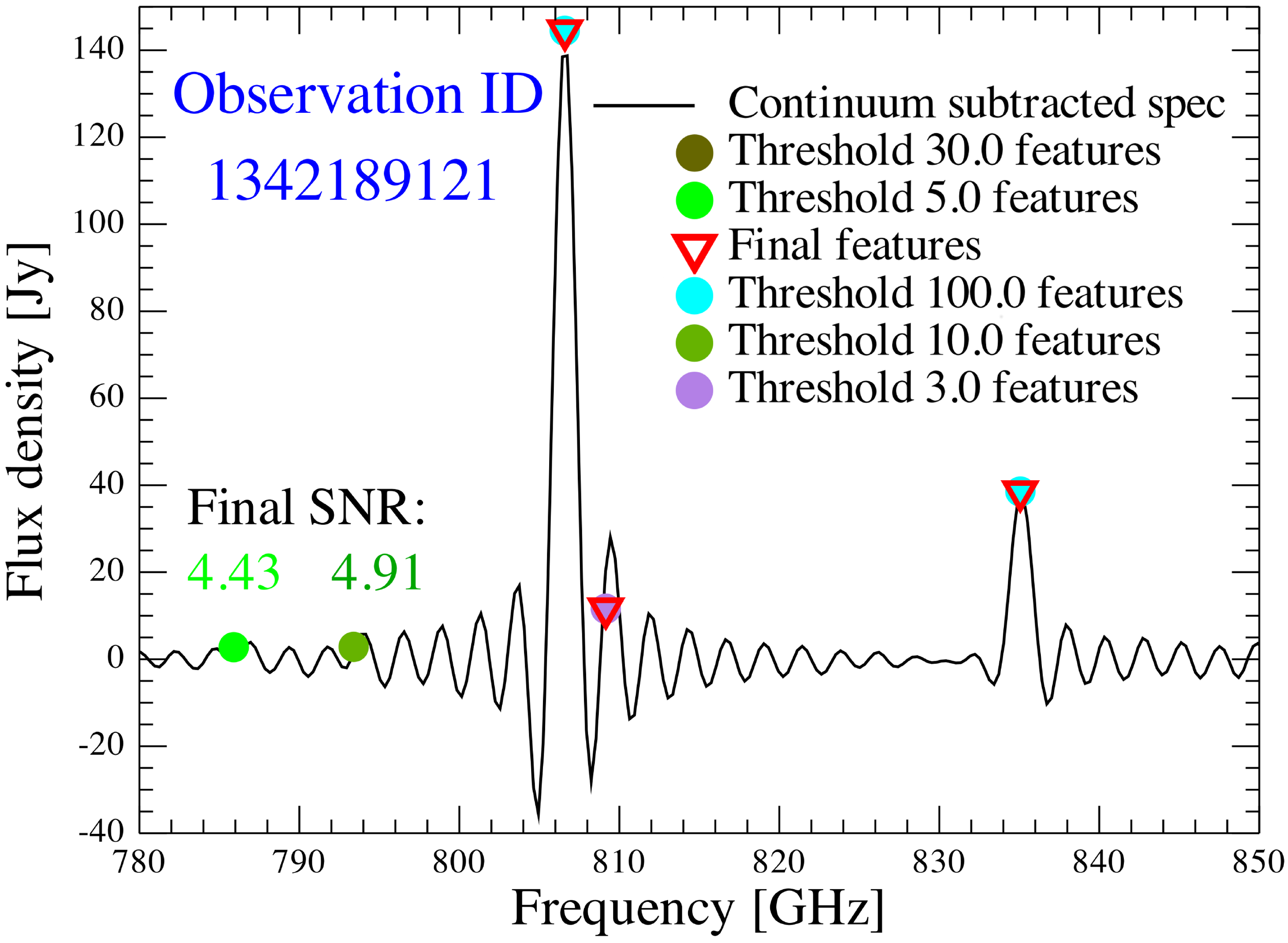}
\includegraphics[trim = 0mm 0mm 0mm 0mm, clip,width=0.33\textwidth]{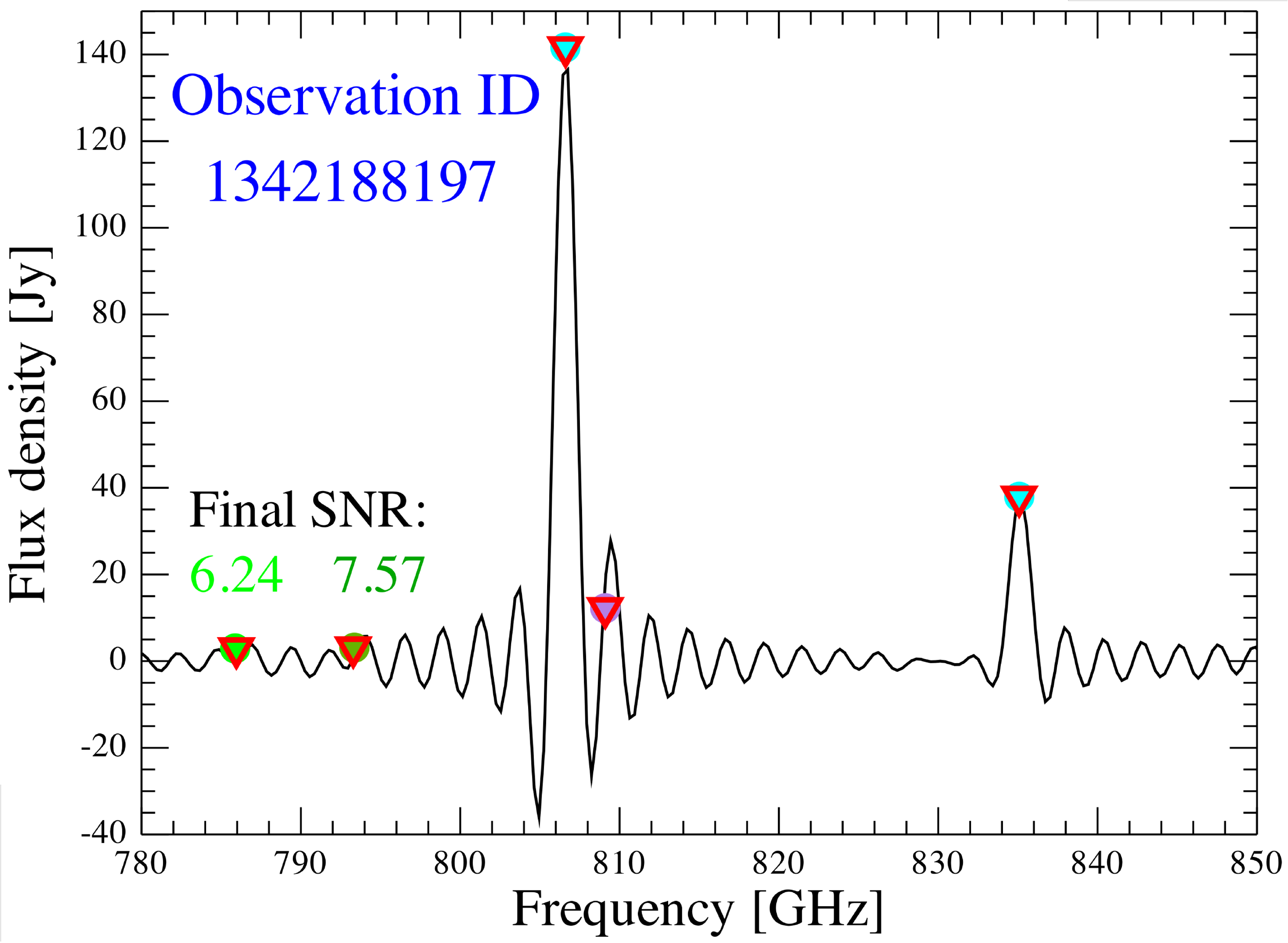}
\includegraphics[trim = 0mm 0mm 0mm 0mm, clip,width=0.33\textwidth]{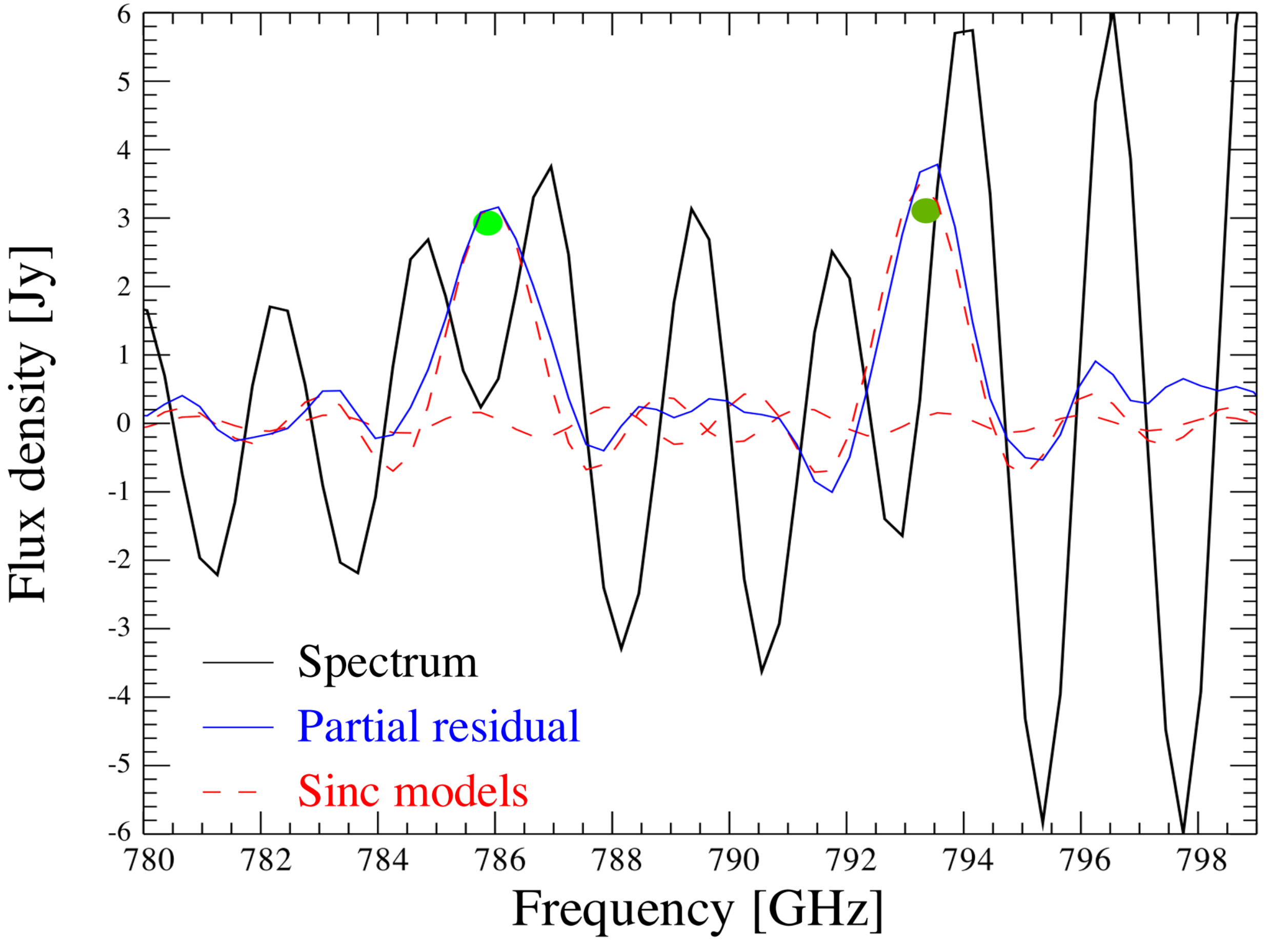}
}
\vspace{-15pt}
\caption{Details of the features found and models fitted for NGC\,7027, around the most blended frequency range ($\sim$ 800\,GHz). The spectrum, all features found, and the features remaining after the final SNR cut are plotted in the left and middle panels, for two observations of NGC\,7027 (i.e., observations 1342189121 and 1342188197). The filled circles show which SNR threshold iteration the features were initially found at, while the red triangles show features in the final catalogue (with a final SNR $>$ 5). On the left, the two lines at``786.0'' and ``793.2''\,GHz (green circles) are just below the final SNR cut, while they are above the SNR cut for the 1342188197 FF catalogue (centre). The right panel shows the residual spectrum for the left case (in blue) after the continuum model and all the sinc models fitted, bar the ones for these two lines, are subtracted. Whereas for the input spectrum (black line) it is hard to see features, there are two clear peaks in the residual at the positions expected.}
\label{fig:ngc7027faint}
\end{figure*}

Fig.\,\ref{fig:ngc7027faint} gives an example of two faint lines (centred at 786.0 and 793.2\,GHz) that the FF catalogues are missing for some observations of NGC\,7027. Although both features are usually found at the 10 or 5 SNR threshold iterations, their final SNRs can fall under the cut of 5 in the final global model fit. The frequency region around these lines ($\sim$780-850\,GHz) also provides a good example of the difficultly of extracting faint features from a spectrum where there are also strong features and significant line blending. Although it should be noted that, relatively speaking, NGC\,7027 does not suffer from high levels of blending. The continuum subtracted SLWC3 spectra from two observations of NGC\,7027 (1342189121 and 1342188197) are plotted in the left and middle panels of Fig.\,\ref{fig:ngc7027faint}, to show the region of interest. In both cases the faint lines are found at the lower positive SNR thresholds, but for 1342189121, these lines have final SNRs $<$ 5. Also in this region is the difficult to detect the [CI]\,$\mathrm{^3P_2 - ^3P_1}$ feature, due to its close proximity to $^{12}$CO(7-6). Both of these features are found for both observations. Looking directly at the spectra, it is hard to see any peaks corresponding to ``786.0'' and ``793.2'\,GHz, while the marker position for [CI]\,$\mathrm{^3P_2 - ^3P_1}$ appears to be horizontally offset from the nearest peak. In the NGC\,7027 panel of Fig.\,\ref{fig:calibratorsHitMiss} there is an insert that shows the spectrum, total model, and sinc profiles from the total fitted model that correspond to [CI]\,$\mathrm{^3P_2 - ^3P_1}$ and $^{12}$CO(7-6). The first positive side-lobe is not well fitted by the sinc for $^{12}$CO(7-6), but after adding the [CI]\,$\mathrm{^3P_2 - ^3P_1}$ sinc, the fit only suffers from the known line asymmetry, leading to the minor bumps seen in the residual. If all the sincs from the total model are subtracted from the input spectrum, bar the two corresponding to ``786.0'' and ``793.2''\,GHz, then the partial residual is that seen in the right panel of Fig.\,\ref{fig:ngc7027faint}, where there are two clear peaks at the expected positions. 

\begin{table}
\begingroup
\begin{center}
\newdimen\tblskip \tblskip=5pt
\caption{\label{tab:rotStats}Percentage of HCO+, HNC and H$_2$O emission lines missed by the FF. The results come from the relevant observations of the SPIRE FTS line calibration sources (detailed in the main text), with the number of observations per species indicated in the table. ``$\overline{\rm SNR}$'' is the mean SNR and ``\% miss'' is the percentage missed for all the rotational transition lines of a particular species that are within the SPIRE FTS frequency range and were measured by \citetalias{Hopwood15} to have SNR $>$ 5. If there is a value in parentheses, this represents the value associated with a difference between the nominal result and the number of times the line was missed when considering all FF features, before the final SNR estimate and cut. Rest frequencies for these transitions are illustrated at the top of Fig.\,\ref{fig:calibratorsHitMiss}.}
\nointerlineskip
\small
%
\newdimen\digitwidth
\setbox0=\hbox{\rm 0}
\digitwidth=\wd0
\catcode`*=\active
\def*{\kern\digitwidth}
\newdimen\signwidth
\setbox0=\hbox{+}
\signwidth=\wd 0
\catcode`!=\active
\def!{\kern\signwidth}
%
\tabskip=2em plus 2em minus 2em
\halign to \hsize{*#\hfil& \hfil#\hfil& \hfil#\hfil& \hfil#\hfil& \hfil#\hfil*\cr
 \multispan5\hrulefill\cr
\noalign{\vspace{-8.0pt}}
 \multispan5\hrulefill\cr
 Species /&      $\nu$&           &                      & \%\cr  
 ***Transition& [GHz]& Detector& $\overline{\rm SNR}$& miss\cr 
\noalign{\vspace{-5.5pt}}
 \multispan5\hrulefill\cr
 HCO+*(31\,observations)& & & & \cr
 ***HCO+(7\,-\,6)*****& *\,624.04& SLWC3& *5.8& 16\,(*0)\cr
 ***HCO+(9\,-\,8)*****& *\,802.52& SLWC3& *6.6& 84\,(77)\cr
\noalign{\vspace{-5.5pt}}
 \multispan5\hrulefill\cr
 
 HNC!*(18\,observations)& & & & \cr
 ***HNC!(*9\,-\,*8)***& *\,815.78& SLWC3& *5.2& *0****\cr
 ***HNC!(11\,-\,10)***& *\,996.86& SLWC3& *5.2& *6\,(*0)\cr
 ***HNC!(11\,-\,10)***& *\,996.86& SSWD4& *5.2& *6\,(*0)\cr
 ***HNC!(12\,-\,11)***& 1\,087.44& SSWD4& *5.3& *6****\cr 
\noalign{\vspace{-5.5pt}}
 \multispan5\hrulefill \cr

  H$_2$O\,!*(18\,observations)& & & & \cr
 ***H$_2$O!\,(2\_02\,-\,1\_11)& *\,988.00& SSWD4& 12.6& 11\,****\cr
 ***H$_2$O!\,(3\_12\,-\,3\_03)& 1\,097.50& SSWD4& 10.4& *6\,****\cr
 ***H$_2$O!\,(1\_11\,-\,0\_00)& 1\,113.42& SSWD4& *5.3& *0\,****\cr
\noalign{\vspace{-7.5pt}}
 \multispan5\hrulefill \cr
\noalign{\vspace{-7.5pt}}
}
\end{center}
\endgroup
\vspace{-12pt}
\end{table}

\begin{figure}
\centering
\includegraphics[trim = 8mm 0mm 8mm 18mm, clip,width=0.98\columnwidth]{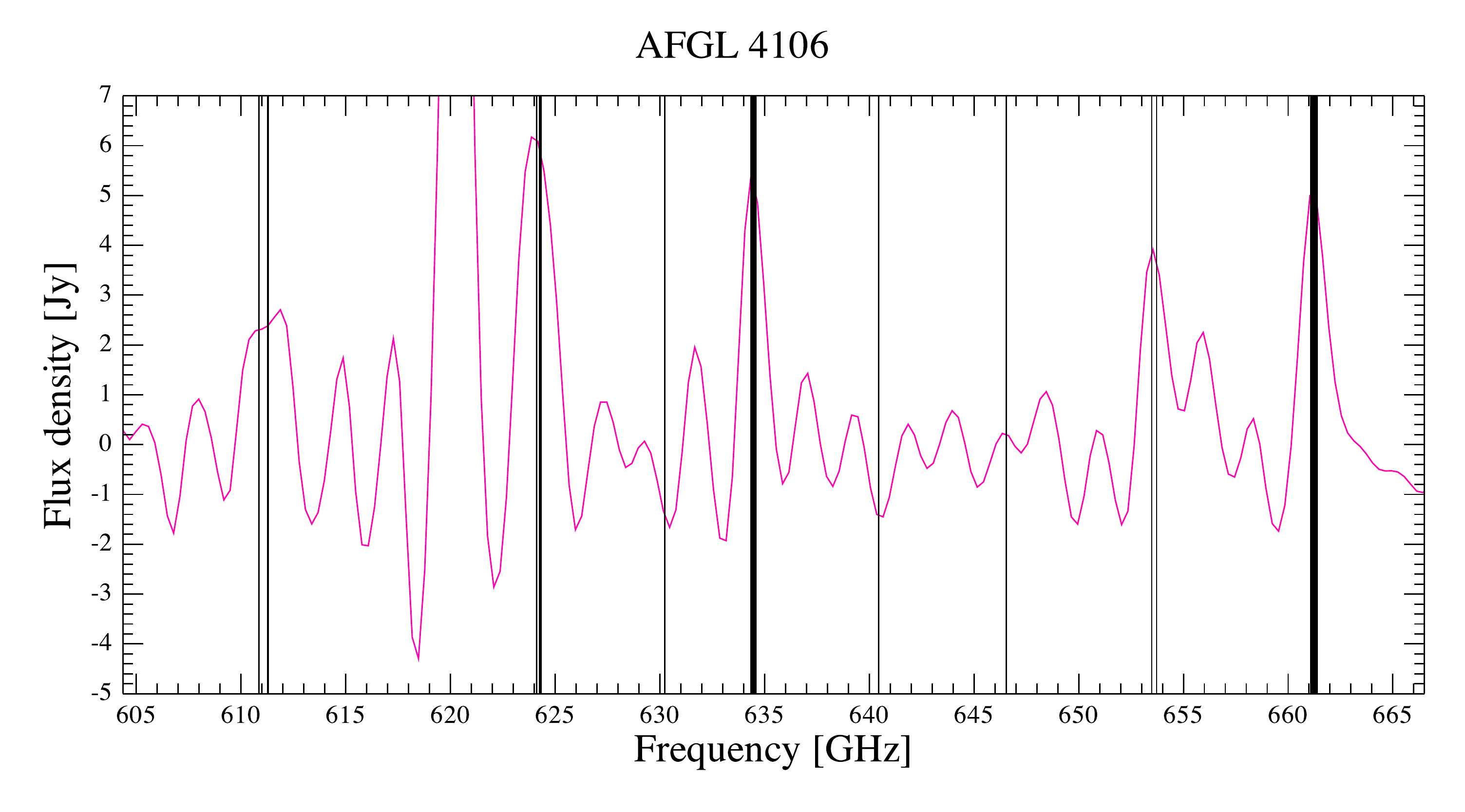}
\vspace{-12pt}
\caption{Example FF features found in observations of AFGL\,4106, which have no match in the \citetalias{Hopwood15} line lists. Thicker lines are caused by the overlaying of a feature at similar frequency in multiple observations of AFGL\,4106. The spectrum plotted is taken from all observations of AFGL\,4106 that have been co-added and continuum subtracted.}
\label{fig:egExtra1}
\vspace{-12pt}
\end{figure}

One last aspect of this comparison are the extra features found. Not all low significance lines were included by \citetalias{Hopwood15}, if these did not improve the overall fit to the data. Therefore some of the extra features can be excluded from the ``likely spurious'' category, as illustrated by Fig.\,\ref{fig:egExtra1}.

\vspace{-12pt}
\subsection{Accessing the products and further information}\label{sec:access}
Full details on the Feature Finder workings and the associated products are available via the main Feature Finder ESA web page\footnote{The \herschel\ SPIRE Spectral Feature Catalogue ESA DOI: \href{https://doi.org/10.5270/esa-lysf2yi}{doi.org/10.5270/esa-lysf2yi}}.
Through the FF pages, one can access documentation, a list of SPIRE FTS literature, and what papers should be referenced when publishing work based on SPIRE FTS data. Tables are provided including all the observations that have an associated FF result. These tables allow access to the FF products and provide a quick way to check the postcards. The combined catalogues and all collections of the postcards can also be accessed. The tables link to FF products stored in the ESA Herschel legacy area\footnote{\label{ff_legacy}The FF \herschel\ legacy area is available at the following URL: \url{http://archives.esac.esa.int/hsa/legacy/HPDP/SPIRE/SPIRE-S/spectral_feature_catalogue/}}. Furthermore, any updates to the FF scripts and products will be documented and provided at this location. Examples of using the FF software, searching the FF catalogues for features of interest, retrieving FF products and HSA data, and using visual inspection and other preliminary analysis tools are provided in a \href{http://archives.esac.esa.int/hsa/legacy/HPDP/SPIRE/SPIRE-S/spectral_feature_catalogue/notebooks/FeatureFinderNotebook_v1.0.html}{Python-based Jupyter notebook}, also available at the legacy FF page\textsuperscript{\ref{ff_legacy}}.
%
\vspace{-18pt}
\section{Summary}\label{sec:summary}
This paper has presented details of the SPIRE FTS Feature Finder process and the associated products. All results are available through the \textit{Herschel}\ Science Archive as Highly Processed Data Products. 
The iterative FF was run on all standard observations of the SPIRE FTS, both in sparse or mapping modes. Line features in each of the two spectral bands were extracted from observations in high resolution mode, while for the low resolution mode we only provide the corresponding parameters of the best fit polynomial continuum. In sparse mode, the two central detectors of each band were considered separately from the off-axis detectors (see \citetalias{FFlineID}), while all spatial pixels were included together for mapping mode. The features are provided with their signal-to-noise ratio: deriving line fluxes is beyond the scope of the automated FF process.

The final combined catalogue of all sparse and mapping observations lists 267\,159 features at $|\mathrm{SNR}| \ge 5$, from 874 (i.e., 669 sparse mode and 205 mapping mode) unique HR observations. The off-axis line search reports an additional 33\,840 spectral features (see \citetalias{FFlineID} for details). Continuum spectra parameters are provided for an additional 199 HR sparse mode observations, and 257 LR sparse mode (only for the central detectors) and 70 LR mapping mode (all spatial pixels) observations.

The FF catalogue is complete for features at SNR above 10, decreasing to 50-70\% completeness down to SNR of 5. The detections were validated with a set of methods and each feature has associated flags that can be used to assess the feature fidelity. There are 7,691 (or 4.6\%) features flagged as a poor fit.

A customised check for the presence of the faint [CI]\,$\mathrm{^3P_2 - ^3P_1}$ line, sometimes buried in the side-lobes of the much stronger $^{12}$CO(7-6) line, leads to an additional 4\,781 features included in the catalogue from 162 unique observations.


Herschel has provided our highest resolution and most sensitive view of the FIR universe to date.
The legacy value of the SPIRE FTS observations will be further enhanced by the spectral feature finder algorithm  presented in this paper. 
\vspace{-30pt}
\section*{Acknowledgments}
\vspace{-3pt}
SPIRE has been developed by a consortium of institutes led by Cardiff Univ. (UK) and including: Univ. Lethbridge (Canada); NAOC (China); CEA, LAM (France); IFSI, Univ. Padua (Italy); IAC (Spain); Stockholm Observatory (Sweden); Imperial College London, RAL, UCL-MSSL, UKATC, Univ. Sussex (UK); and Caltech, JPL, NHSC, Univ. Colorado (USA). This development has been supported by national funding agencies: CSA (Canada); NAOC (China); CEA, CNES, CNRS (France); ASI (Italy); MCINN (Spain); SNSB (Sweden); STFC, UKSA (UK); and NASA (USA). This research is supported in part by the Canadian Space Agency (CSA), the Natural Sciences and Engineering Research Council of Canada (NSERC), Canada Research Chairs (CRC), and CMC Microsystems. This research has made use of the NASA/IPAC Infrared Science Archive, which is operated by the JPL, CalTech, under contract with NASA.\ 
This work used the SciPy (\url{www.scipy.org}) and Astropy (\url{www.astropy.org}) Python packages. Table formatting followed the {\it Planck} Style Guide \citep{PlanckStyle}.
\nocite{2020SciPy-NMeth, astropy:2013, astropy:2018}\\
\vspace{-30pt}
\section*{Data Availability}
\vspace{-3pt}
The \textit{Herschel} SPIRE Spectral Feature Catalogue and code is available at: \href{https://doi.org/10.5270/esa-lysf2yi}{doi.org/10.5270/esa-lysf2yi}.
\vspace{-24pt}
\bibliographystyle{mnras}
\bibliography{v14_spireFtsFeatureFinder}
\label{lastpage}
\end{document}